\documentclass[11pt,fleqn]{article}

\usepackage{graphicx,enumerate}
\usepackage{amsmath}
\usepackage{amssymb}
\usepackage{color}
\usepackage{hyperref}
\hypersetup{colorlinks = true, linktocpage = true, bookmarksopen = true, linkcolor = blue, urlcolor=blue, citecolor = blue, allcolors=blue}
\usepackage[caption=false]{subfig}
\usepackage{authblk}
\usepackage[font=footnotesize]{caption}
\usepackage[numbers,sort&compress]{natbib}
\numberwithin{equation}{section}
\usepackage[margin=2.5cm]{geometry}
\newcommand{\ba}{\begin{array}}
\newcommand{\ea}{\end{array}}

\newcommand{\be}{\begin{equation}}
\newcommand{\ee}{\end{equation}}
\newcommand{\bc}{\begin{center}}
\newcommand{\ec}{\end{center}}
\newcommand{\bdm}{\begin{displaymath}}
\newcommand{\edm}{\end{displaymath}}

\newcommand{\p}{\partial}

\newcommand{\change}[1]{{\color{black}{#1}}}

\date{}

\begin{document}

\title{ \bf
Drug delivery from microcapsules: \\ 
how can we estimate the release time?}

\author[1]{Elliot~J.~Carr}
\author[2]{Giuseppe~Pontrelli}
\affil[1]{{\footnotesize School of Mathematical Sciences, Queensland University of Technology (QUT), Brisbane, Australia}}
\affil[2]{{\footnotesize Istituto per le Applicazioni del Calcolo -- CNR
Via dei Taurini 19 -- 00185 Rome, Italy}}

\maketitle

\begin{abstract}
Predicting the release performance of a drug delivery device is an important challenge in pharmaceutics and biomedical science. In this paper, we consider a multi-layer diffusion model of drug release from a composite spherical microcapsule into an external surrounding medium. Based on this model, we present two approaches that provide useful indicators of the release time, i.e. the time required for the drug-filled capsule to be depleted. Both approaches make use of temporal moments of the drug concentration versus time curve at the centre of the capsule, which provide useful insight into the timescale of the process and can be computed exactly without explicit calculation of the full transient solution of the multi-layer diffusion model. The first approach, which uses the zeroth and first temporal moments only, provides simple algebraic expressions involving the various parameters in the model (e.g. layer diffusivities, mass transfer coefficients, partition coefficients) to characterize the release time while the second approach yields an asymptotic estimate of the release time that depends on consecutive higher moments. Through several test cases, we show that both approaches provide a computationally-cheap and useful measure to compare \textit{a priori} the release time of different composite microcapsule configurations.
\end{abstract}

\vspace*{2ex}\noindent\textit{\bf Keywords}: mass diffusion, drug release, release time, composite capsule, asymptotic estimates.

\section{Introduction}

Polymeric microcapsules are commonly used in pharmaceutical or medical processes as drug carriers or for the encapsulation of organic cells \cite{Larranaga_2017, timin_2016}. The main functions of these \change{micro-} or nano-sized vesicles are the efficient transport and controlled release of the therapeutic agent into the external environment. Typically, drug-filled microcapsules \change{exhibit a multi-layered structure consisting} of a spherical core surrounded by a thin, protective semi-permeable polymeric shell. The release properties depend crucially on the nature of this coating structure \cite{zhang_2013}. \change{For this reason}, in recent years, materials innovation and nanotechnology have stimulated novel research and progress in biodegradable, biocompatible, environment-responsive, and targeted delivery systems \cite{len, cuo}. 
Mathematical modeling plays an important role in elucidating the drug release mechanisms, thus facilitating the development of advanced materials and the assessment of smart
products by a systematic, rather than trial-and-error, approach. \change{Within this area,} special attention is given to semi-empirical and mechanistic models\change{, which} often provide a good fit with experimental data \cite{ari, tavar}. 

\change{Analysis and solution of mechanistic models taking into account the} multi-layer structure of the microcapsule has been recently addressed by \citet{kaoui_2018} and \citet{carr_2018mb}. However, even in situations where the transport phenomena are well represented mathematically, the need exists to develop analytical tools or simple indicators to extract meaning from the model and the data. Such tools make it possible to answer practical questions, such as how long it takes to attain a therapeutic flux, or what processing conditions need to be adjusted and by how much, in order to reach a desired delivery rate without solving the complete mechanistic model. Furthermore, when a full mathematical representation is not available, or is too complicated \change{to be} of any practical use, simple performance indicators are required to elucidate the main transport mechanisms and identify the most critical components of the process. In most circumstances, for example, rather than the full transient solution, the time required to reach a steady state plasma drug concentration \change{solely} determines the effectiveness of the delivery system.

Among several mathematical techniques available for control systems analysis, Laplace transform and linearization techniques are frequently applied to describe a process' dynamics through an effective time constant \cite{collins_1980}: this represents a useful indicator of the time elapsed before reaching a steady state. The concept of a time constant is crucial to pharmaceutical developers who want to create controlled-release devices able to deliver drugs at a desired rate. Estimation of the \change{\emph{release time}, i.e. the time required for the \change{drug-filled} capsule to be depleted}, is of great importance in the design of drug delivery systems, because it allows product manufacturers to tune specific properties to ensure a precise release within a determined time \cite{simon_2009}. The idea of using a single timescale to characterize how fast the drug concentration reaches the equilibrium value has been explored by several authors. \citet{pontrelli_2018} define a timescale equal to the mean of a normalized probability density function representing the transition of the concentration profile from initial to equilibrium state. Calculating this mean requires either computing the full transient solution or evaluating the Laplace transform of the transient solution, and ultimately for multi-layered problems, produces quite complex expressions for the timescale in terms of the relevant parameters in the model. \change{Another approach is the concept of mean action time, which has been used to characterize how long a diffusion process takes to reach steady state \cite{landman_2000,carr_2018pre,simpson_2013,mcnabb_1991}. Here, the transition from initial to equilibrium state is represented as a cumulative distribution function, with the mean of the corresponding probability density function defining the characteristic timescale. The attraction of the mean action time is that it completely avoids any calculation of the transient solution and produces simple explicit algebraic formulas for the timescale in terms of the parameters in the model \cite{carr_2018pre,simpson_2013}. Finally, we note that} the notion of release time \change{is different but related to the concept} of penetration time\change{, which} that has been studied in a number of configurations for thermal disturbances in heat conduction problems \cite{dem2008}. 

In this paper, inspired by recent literature on diffusion processes \cite{carr_2017pre,carr_2018pre,ospi,carr_2018joh,simpson_2013}, \change{we present two approaches that provide useful indicators of the release time for a drug-filled capsule. Both approaches utilize temporal moments of the concentration versus time curve at the centre of the capsule}. The first \change{approach}, which \change{uses the zeroth and} first moments only, provides a simple and cheap way to characterize the release time in terms of the various parameters in the model (e.g. layered diffusivities, mass transfer coefficients, partition coefficients). The second \change{approach yields} an asymptotic estimate of the \change{release time, defined as the time when the \change{drug concentration at the centre of the capsule} is a small prescribed distance away from its steady-state value}. Attractively, both \change{approaches} can be used to avoid explicit (analytical or numerical) solution of the mathematical model and are useful to compare \change{\textit{a priori}} different capsule configurations.

%The attraction of MAT is that it completely avoids any calculation of the transient solution and produces simple explicit algebraic formulas for the timescale in terms of the parameters in the model. Recently, \citet{carr_2018pre} applied the MAT concept to generic layered diffusion problems in Cartesian coordinates deriving simple formulas that provide an indication of the time required for a multi-layer diffusion process to reach equilibrium state. This paper applies, for the first time, the MAT concept to drug delivery problems by extending the above methodology to spherical coordinates thus enabling application to composite spherical microcapsules. 
%The proposed method directly estimates the time constant of linear diffusion problems, either avoiding the explicit (analytical or numerical) solution and the use of Laplace transform. 

The remaining parts of this article are ordered in the following way: (i) the diffusion model for drug release from a spherical multi-layer capsule is presented  (ii) \change{both approaches for providing indicators of the release time are described} (iii) a specific \change{case} study on a core-shell capsule is considered and numerical results presented and discussed.

\section{Drug diffusion model for a multi-layer sphere}
\begin{figure}[t]
\centering
\includegraphics[scale=1.0]{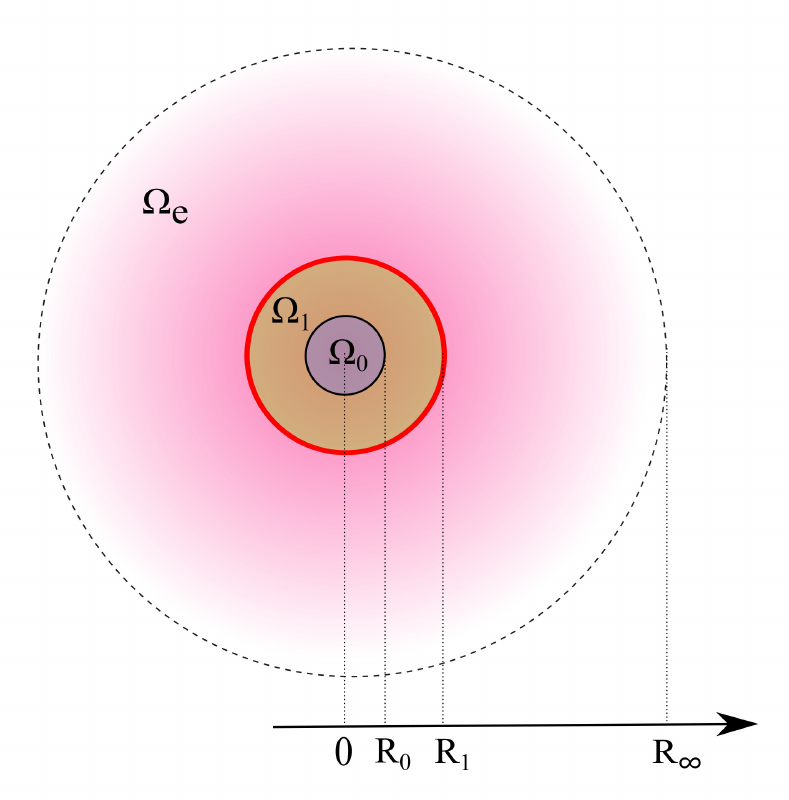}
\caption{Schematic representation of the cross-section of the radially symmetric capsule, comprising an internal core $\Omega_0$ ($0 < r < R_0$), a concentric layer $\Omega_1$ ($R_0 < r < R_1$) and the thin coating shell (in red). This two-layer sphere is immersed in the release medium, represented by a concentric external layer $\Omega_e$, delimited by the dashed line $R_{\infty}$. The length $R_{\infty}$ is named {\em release distance} and is defined as the minimum distance from the sphere surface beyond which  $c(x,t) \simeq 0$, within a prescribed tolerance, at all times. This three enveloped layer system  constitutes  the object of our modelling (figure not to scale).}  
\label{fig:1D_schematic}
\end{figure}

Multi-layer capsules consist of a drug-loaded (fluid or solid) spherical core surrounded by one or more polymeric layers \cite{Larranaga_2017,timin_2016}. Such layer-by-layer assembly enhances a selective diffusion and allows for better control of the transfer rate. In this framework, and in the most general case of $n$ layers, the composite system can be treated as a sequence of enveloping concentric shells, constituted by spheres of increasing radii \change{satisfying} $0 < R_0 < R_1 < \hdots  < R_n$ (see Figure \ref{fig:1D_schematic} \change{for the case $n=1$}). To prevent fast delivery, the capsule's outmost layer is protected with a thin semi-permeable shell ({\em coating}). This coating shields and preserves the encapsulated drug from degradation and chemical aggression, protects the inner structure, and guarantees a more controlled and sustained release \cite{henning_2012}. The coating is in contact with the targeted external release medium (either a bulk fluid or a tissue). Strictly speaking, for a pure diffusion problem into a homogeneous medium, the concentration field undergoes an exponential decay and vanishes asymptotically at infinite distance. In other words, at a given time, the concentration is gradually damped, \change{reaching} zero at infinite distance \cite{cra}. Nevertheless, for computational purposes, we can confine the diffusion process within an enveloping spherical layer of finite extent, at a distance ``far enough'' from the capsule surface. In our model, this additional layer is defined by setting a cut-off length $R_{\infty} \gg R_n $ (sometimes named  {\em release distance} or {\em diffusion length}), beyond which the concentration reduces to zero, within a prescribed tolerance, at all times (Figure \ref{fig:1D_schematic}) \cite{dem2010}:
\bdm
c(x,t) \simeq 0 \qquad \text{for \change{all $x \geq R_{\infty}$ and $t > 0$.}}
\edm

Due to homogeneity and isotropy, we can assume that net drug diffusion occurs along the radial direction only, and thus we restrict our study to a one-dimensional model (Figure \ref{fig:1D_schematic}) that reflects a perfectly radially symmetric system. We adopt the model formulated in \cite{carr_2018mb,kaoui_2018}, where the evolution of concentrations in the layers is governed by a set of \change{one-dimensional} linear diffusion equations:
\begin{alignat}{2}
&{\p c_0 \over \p t} ={D_0 \over r^2} {\p \over \p r}\left(r^2 {\p c_0 \over \p r}\right),&\qquad&\text{$r\in(0, R_0)$,} \label{eq:model_pde1}  \\
&{\p c_i \over \p t} ={D_i \over r^2} {\p \over \p r}\left(r^2 {\p c_i \over \p r}\right),&\qquad&\text{$r\in(R_{i-1} , R_i)$ for $i = 1,\hdots,n$,} \label{eq:model_pde2}  \\
&{\p c_e \over \p t} ={D_e \over r^2} {\p \over \p r}\left(r^2 {\p c_e \over \p r}\right),&\qquad&\text{$r\in(R_{n},R_{\infty})$,} \label{eq:model_pde3}
\end{alignat}
paired with interlayer and boundary conditions
\begin{alignat}{2}
&{\p c_0 \over \p r}=0, &\qquad &\text{$r=0$, } \label{eq:model_bc1}\\
&c_i= \sigma_{i} c_{i+1}, \qquad D_i{ \p c_i \over \p r} = D_{i+1}{ \p c_{i+1} \over \p r},&\qquad&\text{$r=R_i$ for $i = 1,\hdots,n-1$,}\\
&D_n{ \p c_n \over \p r} = D_{e}{ \p c_{e} \over \p r},\qquad D_{e}{ \p c_{e} \over \p r} = P (\sigma_n c_e - c_n), &\quad&\text{$r=R_n$,}\label{eq:model_int2}\\
&c_e = 0, &\qquad& \text{$r=R_{\infty}$.} \label{eq:model_bc2}  
\end{alignat}
In the above equations, the parameters $D_i$ are the diffusion coefficients of drug in each layer, $\sigma_i$ are the partition coefficients, and $P$ is the mass transfer coefficient at the external coating \cite{carr_2018mb,kaoui_2018}. For a releasing drug-loaded core, the initial conditions are:
\begin{alignat}{2}
\label{eq:model_ic1}
&c_0(r,0)=C_0, &\qquad& \text{$r\in(0,R_0)$,}\\
\label{eq:model_ic2}
&c_i(r,0)=0, &\qquad& \text{$r\in(R_{i-1},R_{i})$ for $i = 1,\hdots,n$,}\\
\label{eq:model_ic3}
&c_e(r,0)=0, &\qquad& \text{$r\in(R_{n},R_{\infty})$,}
\end{alignat} 
where $C_0>0$ is a constant. The steady state solution of the drug diffusion model (\ref{eq:model_pde1})--(\ref{eq:model_ic3}) is trivially given by 
\be
\label{eq:model_ss}
c_{i}^{\infty}(r) =\lim_{t \rightarrow \infty} c_{i}(r)= 0, \qquad  \text{for} \quad i = 0,1,\hdots,n,e. 
\ee

In our previous studies the transient solution of Eqs (\ref{eq:model_pde1})--(\ref{eq:model_ic3}) was obtained through a separation of variables \cite{kaoui_2018}, or a Laplace transform approach \cite{carr_2018mb}. In some circumstances, however, rather than working with the complicated expressions of the full solution, simple and cheap measures of the performance of the delivery system are desired. In the following sections, we propose two ways to quantify the \change{\textit{release time}} of the microcapsule\change{, that is, the time taken for the capsule to be depleted of the drug.}

\section{Characterizing the release time}
\label{sec:3}
\begin{figure}[ht]
\centering
\subfloat[(a)]{\includegraphics[height=0.27\textwidth]{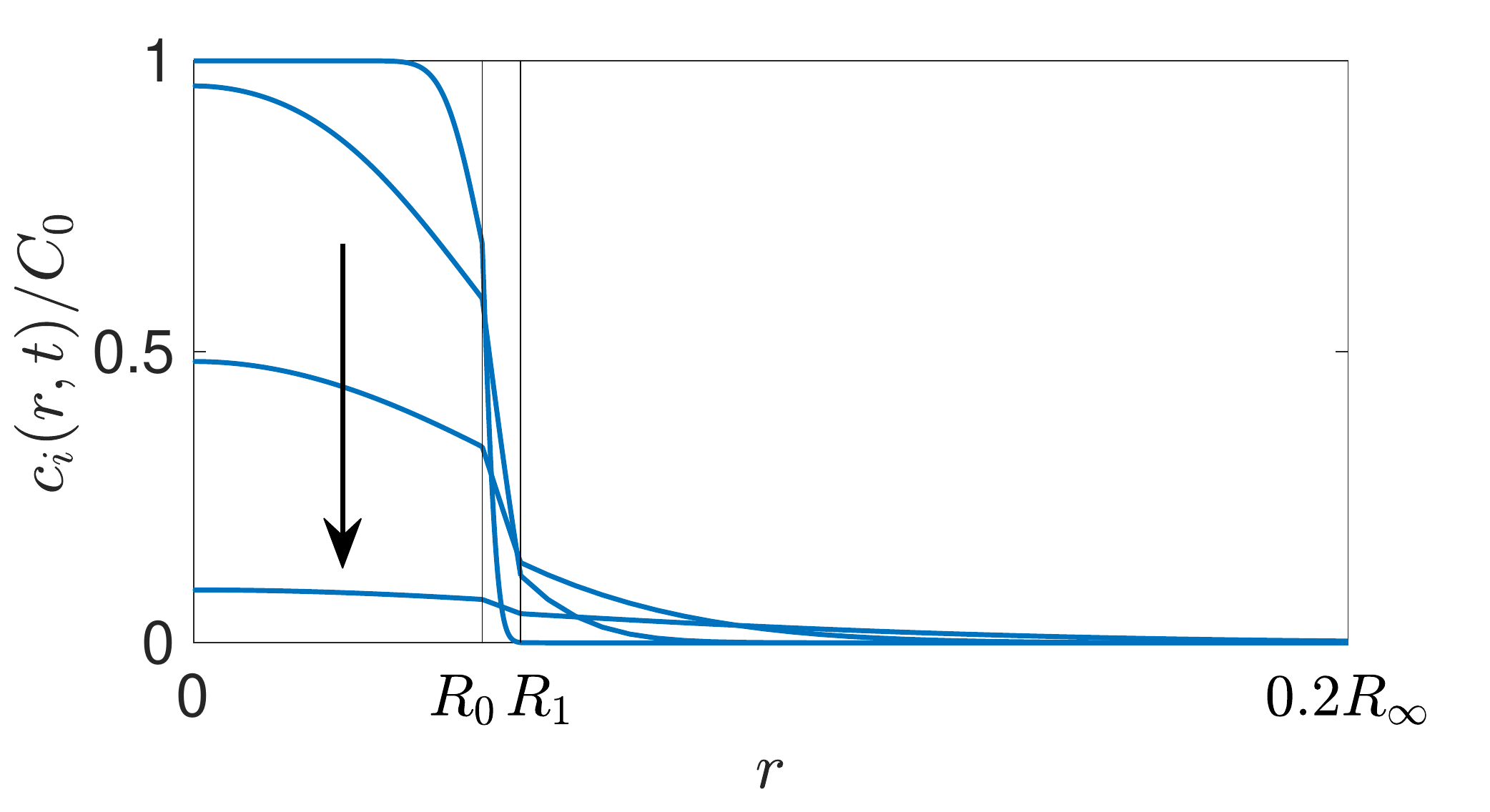}}\hspace{0.0cm}\subfloat[(b)]{\includegraphics[height=0.27\textwidth]{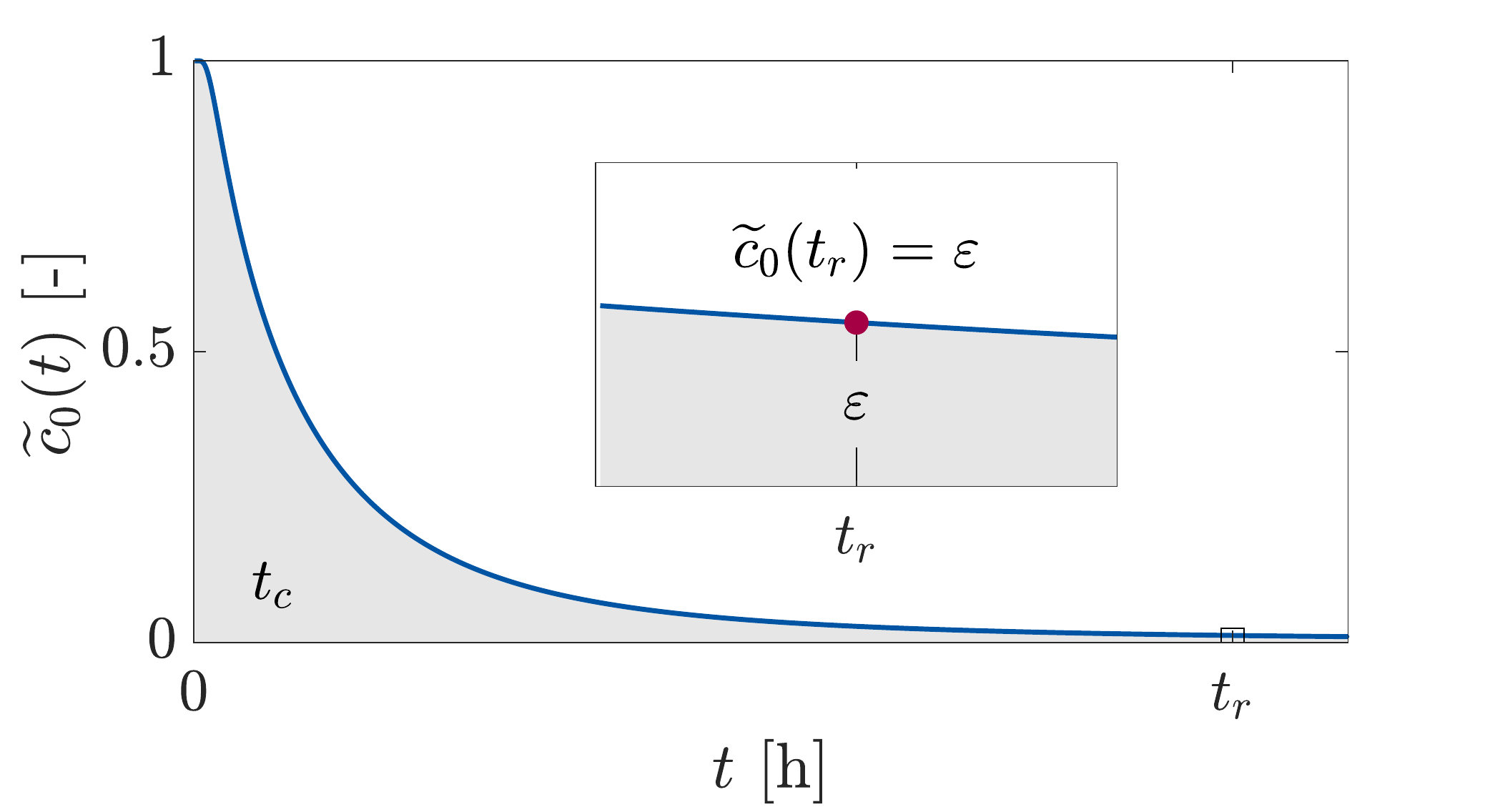}}
\caption{Typical spatio-temporal behaviour of the drug concentration arising from the drug diffusion model (\ref{eq:model_pde1})--(\ref{eq:model_ic3}) with a single hydrogel layer ($n = 1$) (a) Plot of the dimensionless concentration ($c_{i}(r,t)/C_{0}$) versus radius ($r$) at \change{four values of $t$} with a black arrow indicating the direction of increasing time. Vertical lines denote the interfaces $r = R_{0}$ and $r = R_{1}$. (b) Plot of the dimensionless concentration at the centre of the capsule $\widetilde{c}_{0}(t)$ versus time. The area underneath this curve defines the characteristic timescale $t_{c}^{(1)}$ (\ref{eq:tc1}) while the release time\change{, as defined in Section \ref{sec:4}, is} the time when $\widetilde{c}_{0}(t)$ reaches a small prescribed value $\varepsilon$.}
\label{fig:Concentration_example}
\end{figure}

%From this curve, we define the {\em release time} 
%\underline{Characterizing the release time} \\
%{\bf We now present an alternative indicator of the release timescale based on its asymptotic decay of the concentration $\widetilde{c}_{0}(t)$, making use of its temporal moment (REFERENCE) that gets rid of the arbitrary accuracy measure $\varepsilon$.  	\par
%
%While the normalized concentration $\widetilde{c}_{0}(t)$ is always dropping down between 1 and 0, 
%the shape of this function is a good indicator of the character of the release.}
\change{In this section, we define two indicators that provide a useful characterisation of the release time scale.} Figure \ref{fig:Concentration_example} shows the typical spatio-temporal behaviour of the dimensionless concentration, $c_{i}(r,t)/C_{0}$, arising from the solution of the diffusion model (\ref{eq:model_pde1})--(\ref{eq:model_ic3}). An important observation from this plot is that the centre of the capsule, $r=0$, takes the longest time to reach steady state. In Figure \ref{fig:Concentration_example}b we plot the dimensionless concentration at the centre of the capsule, $\widetilde{c}_{0}(t) := c_{0}(0,t)/C_{0}$, versus time as it progresses towards the steady state solution of zero concentration (\ref{eq:model_ss}). The area underneath the curve $\widetilde{c}_{0}(t)$, or the zeroth temporal moment of $\widetilde{c}_{0}(t)$, is defined as
\begin{gather}
\label{eq:tc1}
t_{c}^{(1)} = \int_{0}^{\infty} \widetilde{c}_{0}(t)\,\mathrm{d}t,
\end{gather}
and shaded in Figure \ref{fig:Concentration_example}b. \change{The value of $t_{c}^{(1)}$} will tend \change{to be small} for a fast release and \change{large} for a slow release. Moreover, when comparing two different capsule configurations, if the corresponding $\widetilde{c}_{0}(t)$ curves for the two configurations do not intersect then the configuration with the larger release time will have a larger area (larger value of $t_{c}^{(1)}$). Therefore, it is reasonable to conclude that $t_{c}^{(1)}$ provides a useful characterization of the timescale of release\footnote{\change{We remark that Eq (\ref{eq:tc1}) differs from the standard definition of mean action time (see e.g.~\cite{carr_2018pre}) commonly used to characterize the time taken for a diffusion process to reach steady state. The reason for this difference is that the standard definition is not defined when $c_{i}(r,0) = c_{i}^{\infty}(r)$, as is the case in Eqs (\ref{eq:model_ic1})--(\ref{eq:model_ss}).}}.  

A major attraction of working with Eq (\ref{eq:tc1}) is that a simple closed-form expression can be derived for $t_{c}^{(1)}$ involving the model parameters without requiring the full expression for $c_{0}(0,t)$. \change{This is achieved by extending and modifying similar ideas presented elsewhere (see, e.g., \cite{carr_2018pre,simpson_2013})}. First, we define:
\begin{gather}
\label{eq:ui}
u_{i}(r) = \frac{1}{C_{0}}\int_{0}^{\infty}c_{i}(r,t)\,\mathrm{d}t,
\end{gather}
which allow us to write down an equivalent form of Eq (\ref{eq:tc1})
\begin{gather}
\label{eq:tc2}
t_{c}^{(1)} = u_{0}(0).
\end{gather}
Next, applying the linear operator $\mathcal{L}$, defined as
\begin{gather}
\label{eq:linear_operator}
\mathcal{L}\varphi := \frac{D_{i}}{r^{2}}\frac{\textrm{d}}{\textrm{d}r}\left(r^{2}\frac{\textrm{d}\varphi}{\textrm{d}r}\right),
\end{gather}
to both sides of Eq (\ref{eq:ui}) and making use of Eqs (\ref{eq:model_pde1})--(\ref{eq:model_pde3}) yields the following differential equation:
\begin{align}
\label{eq:mat_ode}
{D_{i} \over r^2}\frac{\textrm{d}}{\textrm{d}r}\left(r^2 \frac{\textrm{d}u_{i}}{\textrm{d}r}\right) &= \frac{1}{C_{0}}\bigl[c_{i}^{\infty}(r) - c_{i}(r,0)\bigr].
\end{align}
Supplementary boundary and interlayer conditions corresponding to Eqs (\ref{eq:model_bc1})--(\ref{eq:model_bc2}) are derived using the definition of $u_{i}(r)$ in Eq (\ref{eq:ui}) (see, e.g. \cite{carr_2017pre, carr_2018pre}). In summary, recalling the initial conditions (\ref{eq:model_ic1})--(\ref{eq:model_ic3}) and the steady-state solution (\ref{eq:model_ss}), we have the following boundary value problem satisfied by $u_{i}(r)$ ($i = 0,1,\hdots,n,e$):
\begin{alignat}{2}
&{D_0 \over r^2} {\textrm{d}\over \textrm{d}r}\left(r^2 {\textrm{d} u_0 \over \textrm{d} r}\right) = -1, &\qquad&\text{$r\in(0 , R_0)$} \label{eq:bvp_ode1}\\
&{D_i \over r^2} {\textrm{d}\over \textrm{d}r}\left(r^2 {\textrm{d} u_i \over \textrm{d} r}\right) = 0, &\qquad&\text{$r\in(R_{i-1} , R_i)$ for $i = 1,\hdots,n$,} \label{eq:bvp_ode2}\\
&{D_e \over r^2} {\textrm{d} \over \textrm{d} r}\left(r^2 {\textrm{d} u_e \over \textrm{d} r}\right) = 0, &\qquad&\text{$r \in( R_{n}, R_{\infty} ) $,} \label{eq:bvp_ode3}  \\
&{\textrm{d} u_0 \over \textrm{d} r}=0,&\quad &\text{$r=0$,} \label{eq:bvp_bc1}\\
&u_i= \sigma_{i} u_{i+1}, \quad
D_i{ \textrm{d} u_i \over \textrm{d} r} = D_{i+1}{ \textrm{d} u_{i+1} \over \textrm{d} r}, &\qquad& \text{$r=R_i$ for $i = 0,1,\hdots,n-1$,} \label{eq:bvp_int1}\\
&D_n{ \textrm{d} u_n \over \textrm{d} r} = D_{e}{ \textrm{d} u_{e} \over \textrm{d} r}, \quad D_{e}{ \textrm{d} u_{e} \over \textrm{d} r} = P (\sigma_n u_e - u_n), &\quad&\text{$r=R_n$,}\label{eq:bvp_int2}\\
&u_e = 0, &\qquad& \text{$r = R_{\infty}$.} \label{eq:bvp_bc2}  
\end{alignat}
The above boundary value problem admits a closed-form analytical solution. By way of example, we consider the simplest case of the \textit{core-shell model}, a drug-filled core surrounded by one hydrogel shell ($n=1$). In this case, the differential equations (\ref{eq:bvp_ode1})--(\ref{eq:bvp_ode3}) possess the general solution:
\begin{align}
\label{eq:u0}
& u_0(r)= {\alpha_0 \over r} + \alpha_1 - \frac{r^{2}}{6D_{0}},\\
\label{eq:u1}
& u_1(r)= {\alpha_2 \over r} + \alpha_3,\\
\label{eq:ue}
& u_e(r)= {\alpha_4 \over r} + \alpha_5, 
\end{align}
where $\alpha_0, \alpha_{1}, \hdots, \alpha_5$ are arbitrary constants. Immediately, the boundary condition (\ref{eq:bvp_bc1}) requires $\alpha_{0} = 0$. The remaining constants satisfy the following algebraic system generated by substituting Eqs (\ref{eq:u0})--(\ref{eq:ue}) into the four interface conditions (\ref{eq:bvp_int1})--(\ref{eq:bvp_int2}) with $n = 1$:
\begin{align}
\label{eq:system1}
&\alpha_{1} - \frac{R_{0}^{2}}{6D_{0}} = \sigma_0 \left({\alpha_2 \over R_0} + \alpha_3 \right),\\
\label{eq:system2}
&{R_{0} \over 3} = {D_1 \alpha_2 \over R_0^2 },\\
\label{eq:system3}
&{D_{1} \alpha_2  \over R_1^2} = {D_{e} \alpha_4 \over R_1^2},\\
\label{eq:system4}
&-{D_{e} \alpha_4  \over R_1^2P} = \sigma_{1}\left(\frac{\alpha_{4}}{R_{1}} + \alpha_{5}\right) - \frac{\alpha_{2}}{R_{1}}-\alpha_{3},\\
\label{eq:system5}
&\frac{\alpha_{4}}{R_{\infty}} + \alpha_{5} =0.
\end{align}
Eqs (\ref{eq:system1})--(\ref{eq:system5}) can then be solved sequentially: Eq (\ref{eq:system2}) for $\alpha_{2}$, Eq (\ref{eq:system3}) for $\alpha_4$, Eq (\ref{eq:system5}) for $\alpha_{5}$, Eq (\ref{eq:system4}) for $\alpha_3$ and finally Eq (\ref{eq:system1}) for $\alpha_{1}$. Combining these results with Eqs (\ref{eq:u0})--(\ref{eq:ue}) yields the solution to the boundary value problem (\ref{eq:bvp_ode1})--(\ref{eq:bvp_bc2})\footnote{\change{Note that evaluating either $u_{0}(r)$, $u_{1}(r)$ or $u_{e}(r)$ allows the area under the $\widetilde{c}_{0}(t)$ curve between $t = 0$ and $t\rightarrow\infty$ to be calculated for any value of $r$, which may also be of practical interest.}}:
\begin{alignat}{2}
&u_{0}(r) = \frac{R_{0}^{2}-r^{2}}{6D_{0}}
 + \frac{\sigma_{0}R_{0}^2(R_{1}-R_{0})}{3D_{1}R_{1}} 
+ \frac{\sigma_{0}\sigma_{1}R_{0}^{3}(R_{\infty}-R_{1})}{3D_{e}R_{1}R_{\infty}} + \frac{\sigma_{0}R_{0}^{3}}{3R_{1}^{2}P}, &\quad& r\in (0,R_{0}),\label{eq:u0_final}\\
&u_{1}(r) = \frac{R_{0}^{3}(r-R_{1})}{3D_{1}R_{1}r} + \frac{\sigma_{1}R_{0}^{3}(R_{\infty}-R_{1})}{3D_{e}R_{1}R_{\infty}} + \frac{R_{0}^{3}}{3R_{1}^{2}P}, && r\in (R_{0},R_{1}),
\label{eq:ue_final} \\
&u_{e}(r) = \frac{R_{0}^{3}(R_{\infty}-r)}{3D_{e}R_{\infty}r}, && r \in (R_{1},R_{\infty}) \label{eq:ue1_final}.
\end{alignat}
Evaluating Eq (\ref{eq:u0_final}) at $r = 0$ gives the following expression for the characteristic timescale (\ref{eq:tc2}):
\begin{align}
\label{eq:tc1final}
t_{c}^{(1)} = \frac{R_0^2}{6D_{0}} + \frac{\sigma_{0}R_{0}^{2}\left(R_{1} - R_{0}\right)}{3D_{1}R_{1}} + \frac{\sigma_{0}\sigma_{1}R_{0}^{3}\left(R_{\infty}-R_{1}\right)}{3D_{e}R_{1}R_{\infty}} + \frac{\sigma_{0}R_{0}^{3}}{3R_{1}^{2}P}.
\end{align}
In summary, Eq (\ref{eq:tc1final}) provides a simple formula for characterizing the timescale of release: for small values of $t_{c}^{(1)}$ we expect a rapid release while for large values of $t_{c}^{(1)}$ we expect a slow release \cite{carr_2018pre}.

An alternative to $t_{c}^{(1)}$ is to incorporate the first temporal moment, $\int_{0}^{\infty} t\widetilde{c}_{0}(t)\,\mathrm{d}t$, into the calculation \cite{simpson_2013}:
\begin{gather}
\label{eq:tc2bis}
t_{c}^{(2)} = \int_{0}^{\infty} \widetilde{c}_{0}(t)\,\mathrm{d}t + \sqrt{\int_{0}^{\infty} t\widetilde{c}_{0}(t)\,\mathrm{d}t} = t_{c}^{(1)} + \sqrt{\int_{0}^{\infty} t\widetilde{c}_{0}(t)\,\mathrm{d}t},
\end{gather}
where the square root ensures that the second term has units of time. Note that $t_{c}^{(2)} > t_{c}^{(1)}$ with the inclusion of this additional term in Eq (\ref{eq:tc2bis}) introducing a penalty to $\widetilde{c}_{0}(t)$ curves with heavy tails \change{that exhibit} a slower decay to zero, \change{behaviour that may not be captured by the zeroth moment alone}. By deriving a similar boundary value problem to Eqs (\ref{eq:bvp_ode1})--(\ref{eq:bvp_bc2}) \change{for the first moment}, as described later in Section \ref{sec:4}, the following closed-form expression can be derived:
\begin{align}
&t_{c}^{(2)} = \frac{R_0^2}{6D_{0}} + \frac{\sigma_{0}R_{0}^{2}\left(R_{1} - R_{0}\right)}{3D_{1}R_{1}} + \frac{\sigma_{0}\sigma_{1}R_{0}^{3}\left(R_{\infty}-R_{1}\right)}{3D_{e}R_{1}R_{\infty}} + \frac{\sigma_{0}R_{0}^{3}}{3R_{1}^{2}P}\nonumber\\ 
&\hphantom{t_{c}^{(2)} = \frac{R_0^2}{6D_{0}}} + \left\{\frac{7R_{0}^{4}}{180D_{0}^{2}} + \frac{7\sigma_{0}R_{0}^{4}(R_{1}-R_{0})}{45D_{0}D_{1}R_{1}} + \frac{7\sigma_{0}\sigma_{1}R_{0}^{5}(R_{\infty}-R_{1})}{45D_{0}D_{e}R_{1}R_{\infty}}+ \frac{7\sigma_{0}R_{0}^{5}}{45D_{0}PR_{1}^{2}}\right.\nonumber\\
&\hphantom{t_{c}^{(2)} = \frac{R_0^2}{6D_{0}}}+ \frac{2\sigma_{0}R_{0}^{3}(R_{1}-R_{0})^{2}(\sigma_{0}R_{0} + R_{1}-R_{0})}{9D_{1}^{2}R_{1}^{2}} + \frac{4\sigma_{0}\sigma_{1}R_{0}^{3}(R_{\infty}-R_{1})(\sigma_{0}R_{0}^{3}-R_{0}^{3}+R_{1}^{3})}{9D_{e}PR_{1}^{3}R_{\infty}}\nonumber\\
&\hphantom{t_{c}^{(2)} = \frac{R_0^2}{6D_{0}}}+ \frac{2\sigma_{0}R_{0}^{3}(R_{1}-R_{0})(2\sigma_{0}R_{0}^{2} - 2R_{0}^{2} + R_{0}R_{1}+R_{1}^{2})}{9D_{1}R_{1}^{2}}\left[\frac{\sigma_{1}(R_{\infty}-R_{1})}{D_{e}R_{\infty}} + \frac{1}{PR_{1}}\right]\nonumber\\
&\hphantom{t_{c}^{(2)} = \frac{R_0^2}{6D_{0}}}\left. + \frac{2\sigma_{0}\sigma_{1}R_{0}^{3}(R_{\infty}-R_{1})^{2}(\sigma_{0}\sigma_{1}R_{0}^{3} - \sigma_{1}R_{0}^{3} + \sigma_{1}R_{1}^{3} - R_{1}^{3} + R_{1}^{2}R_{\infty})}{9D_{e}^{2}R_{1}^{2}R_{\infty}^{2}}\right\}^{1/2}.\label{eq:tc2final}
\end{align}
The advantage of \change{incorporating information about} the first moment is that the expression (\ref{eq:tc2final}), although more complex and costly than (\ref{eq:tc1final}), provides a better indication of the release time (see section \ref{sec:results} for additional comments on the results). Other time scale indicators can be similarly defined by the natural extension of (\ref{eq:tc2bis}) to higher moments, but because of their increasingly complicated form, they \change{likely have} limited practical use.

\section{Estimating the release time using high order moments} 
\label{sec:4}
The characteristic timescales, $t_{c}^{(1)}$ (\ref{eq:tc1final}) or $t_{c}^{(2)}$ (\ref{eq:tc2final}), provide an \textit{indicator} for, not an estimate of, the release time of the capsule. In this section, we present an asymptotic estimate of the release time 
defined as the time $t_{r}>0$ satisfying:
\begin{gather}
\widetilde{c}_{0}(t_{r}) = \varepsilon, \label{eq:tr_definition}
\end{gather}
where $\varepsilon$ is a small specified tolerance (see Figure \ref{fig:Concentration_example}b). \change{In other words, $t_r$ measures the time taken for the capsule to be depleted  within a small tolerance.} Since $\widetilde{c}_{0}(t)$ monotonically decreases from one to zero, $t_{r}$ is unique \change{for a given choice of $\varepsilon$. Under this definition,} the release time is not an absolute concept, but \change{dependent on the} desired level of accuracy required to measure a complete depletion. 

\change{To derive the asymptotic estimate of the release time,} we extend to multi-layer diffusion in spherical coordinates previous work by \citet{carr_2017pre}, which focussed on monolayer diffusion in Cartesian coordinates. At $r = 0$, the analytical solution to the drug diffusion model (\ref{eq:model_pde1})--(\ref{eq:model_ic3}) possesses the following functional form \cite{kaoui_2018}:
\begin{gather}
\label{eq:ci}
\widetilde{c}_{0}(t) = \sum_{j=0}^{\infty}\gamma_{j}e^{-t\beta_{j}},\quad 0 < \beta_{0} < \beta_{1} < \beta_{2} < \hdots,
\end{gather}
where $\gamma_{j}$ and $\beta_{j}$ are constants. It follows then that the long time behaviour of the concentration is exponentially decreasing:
\begin{gather}
\label{eq:ci_asymptotic}
\widetilde{c}_{0}(t) \sim \gamma_{0} e^{-t\beta_{0}},\quad\text{as $t\rightarrow\infty$}.
\end{gather}
Combining Eqs (\ref{eq:tr_definition}) and (\ref{eq:ci_asymptotic}) gives the following asymptotic estimate of the release time\footnote{\change{To guarantee $t_{r}>0$, the condition $\varepsilon < \gamma_{0}$  is required.}}:
\begin{align}
\label{eq:tr}
t_{r} \sim \frac{\ln(\gamma_{0}/\varepsilon)}{\beta_{0}}.
\end{align} 

The formula (\ref{eq:tr}) requires knowledge of $\gamma_{0}$ and $\beta_{0}$, which are related to the dominant eigenvalue and eigenfunction pair of the underlying Sturm-Liouville problem \cite{kaoui_2018}. Alternatively, $\gamma_{0}$ and $\beta_{0}$ can be calculated by \change{using} appropriate temporal moments of the concentration as we now describe. Define the $k$th temporal moment\footnote{Under this definition the functions $u_{i}(r)$ defined in Eq (\ref{eq:ui}) can be thought as the zeroth temporal moment, i.e., $u_{i,k}(r)$ with $k=0$.}:
\begin{gather}
\label{eq:kmoment}
u_{i,k}(r) = \frac{1}{C_{0}}\int_{0}^{\infty} t^{k}c_{i}(r,t)\,\text{d}t,
\end{gather}
where the first subscript indicates the layer ($i=0,1,...,e$) and the second subscript denotes the $k$th moment ($k=0,1,\hdots$). Combining Eqs (\ref{eq:ci}) and (\ref{eq:kmoment}) and carrying out the integration yields:
\begin{gather}
u_{0,k}(0) = \sum_{j=0}^{\infty} \frac{k!\gamma_{j}}{\beta_{j}^{k+1}}.
\end{gather}
Since $\beta_{0} < \beta_{j}$ for all $j = 1,2,\hdots$, we have the following asymptotic relation for the higher moments \cite{carr_2017pre}:
\begin{gather}
\label{eq:kmoment_asymptotic}
u_{0,k}(0) \sim \frac{k!\,\gamma_{0}}{\beta_{0}^{k+1}},\quad\text{as $k\rightarrow\infty$}.
\end{gather}
Evaluating Eq (\ref{eq:kmoment_asymptotic}) at $k$ and $k-1$ and solving the resulting algebraic system for $\gamma_{0}$ and $\beta_{0}$ gives:
\begin{gather}
\label{eq:alpha_beta}
\gamma_{0} = \frac{u_{0,k}(0)}{k!}\left(\frac{k \,u_{0,k-1}(0)}{u_{0,k}(0)}\right)^{k+1},\quad
\beta_{0} = \frac{k \, u_{0,k-1}(0)}{u_{0,k}(0)}.
\end{gather}
Finally, substituting Eq (\ref{eq:alpha_beta}) into Eq (\ref{eq:tr}) yields the following asymptotic estimate of the release time:
\begin{align}
\label{eq:tr_estimate}
t_{r} \sim \frac{u_{0,k}(0)}{ku_{0,k-1}(0)}\ln \left[\frac{u_{0,k}(0)}{k!\,\varepsilon}\left(\frac{ku_{0,k-1}(0)}{u_{0,k}(0)}\right)^{k}\right] =: t_{r}^{\ast},
\end{align}
where the superscript ($\ast$) is used to signify that $t_{r}^{\ast}$ is an asymptotic estimate of $t_{r}$. Due to Eq (\ref{eq:kmoment_asymptotic}), the expectation is that $t_{r}^{\ast}$ becomes more accurate as $k$ increases \cite{carr_2017pre}. The attraction of the formula for $t_{r}^{\ast}$ is that the moment expressions appearing in Eq (\ref{eq:tr_estimate}) can be calculated without computing the full transient solution $c_{i}(r,t)$, which appears in the definition (\ref{eq:kmoment}). This is achieved by deriving a similar boundary value problem to the one satisfied by $u_{i}(r)$ (or, equivalently $u_{i,0}(r)$) given in Eqs (\ref{eq:bvp_ode1})--(\ref{eq:bvp_bc2}). Applying the linear operator $\mathcal{L}$ in Eq (\ref{eq:linear_operator}) to Eq (\ref{eq:kmoment}) and utilising Eqs (\ref{eq:model_pde1})--(\ref{eq:model_pde3}) yields:
\begin{align}
&{D_i \over r^2} {\textrm{d}\over \textrm{d}r}\left(r^2 {\textrm{d} u_{i,k} \over \textrm{d} r}\right) = \frac{1}{C_{0}}\int_{0}^{\infty} t^{k}\frac{\partial c_{i}}{\partial t}\,\text{d}t.
\end{align}
Integrating by parts and noting that $c_{i}(r,t)\rightarrow 0$ exponentially as $t\rightarrow\infty$ produces the following differential equation for $u_{i,k-1}(r)$:
\begin{align}
&{D_i \over r^2} {\textrm{d}\over \textrm{d}r}\left(r^2 {\textrm{d} u_{i,k} \over \textrm{d} r}\right) = -ku_{i,k-1},
\end{align}
involving the $(k-1)$th moment, $u_{i,k-1}(r)$. Similar boundary and interlayer conditions to those in Eqs (\ref{eq:bvp_ode1})--(\ref{eq:bvp_bc2}) apply \cite{carr_2018pre} giving the following boundary value problem for the $k$th moment:
\begin{alignat}{2}
\label{eq:kbvp_ode1}
&{D_0 \over r^2} {\textrm{d}\over \textrm{d}r}\left(r^2 {\textrm{d} u_{0,k} \over \textrm{d} r}\right) = -ku_{0,k-1}, &\quad&\text{$r\in(0 , R_0)$}\\
\label{eq:kbvp_ode2}
&{D_i \over r^2} {\textrm{d}\over \textrm{d}r}\left(r^2 {\textrm{d} u_{i,k} \over \textrm{d} r}\right) = -ku_{i,k-1}, &&\text{$r\in(R_{i-1} , R_i)$, $i = 1,\hdots,n$,}\\
\label{eq:kbvp_ode3}
&{D_e \over r^2} {\textrm{d} \over \textrm{d} r}\left(r^2 {\textrm{d} u_{e,k} \over \textrm{d} r}\right) = -ku_{e,k-1}, &&\text{$r \in(R_{n}, R_{\infty})$,}\\
&{\textrm{d} u_{0,k} \over \textrm{d} r}=0,&\quad &\text{$r=0$,} \label{eq:kbvp_bc1}\\
&u_{i,k}= \sigma_{i} u_{i+1,k}, \quad
D_i{ \textrm{d} u_{i,k} \over \textrm{d} r} = D_{i+1}{ \textrm{d} u_{i+1,k} \over \textrm{d} r}, && \text{$r=R_i$, $i = 0,1,\hdots,n-1$,} \label{eq:kbvp_int1}\\
&D_n{ \textrm{d} u_{n,k} \over \textrm{d} r} = D_{e}{ \textrm{d} u_{e,k} \over \textrm{d} r}, \quad D_{e}{ \textrm{d} u_{e,k} \over \textrm{d} r} = P (\sigma_n u_{e,k} - u_{n,k}), &\qquad&\text{$r=R_n$,}\label{eq:kbvp_int2}\\
&u_{e,k} = 0, && \text{$r = R_{\infty}$.} \label{eq:kbvp_bc2} 
\end{alignat}
The release time is computed iteratively by solving the sequence of boundary value problems (\ref{eq:kbvp_ode1})--(\ref{eq:kbvp_bc2}) for increasing values of $k$ \cite{carr_2017pre,carr_2018joh}. First, an initial estimate of $t_{r}^{\ast}$ is calculated using $k = 1$ in Eq (\ref{eq:tr_estimate}) by solving Eqs (\ref{eq:kbvp_ode1})--(\ref{eq:kbvp_bc2}) with $k = 1$ and by using the zeroth order moment $u_{i,0}(r)$ computed previously (Section \ref{sec:3}) \change{in the right-hand side of Eqs (\ref{eq:kbvp_ode1})--(\ref{eq:kbvp_ode3})}. Next, an improved estimate of $t_{r}^{\ast}$   is computed using $k = 2$ in Eq (\ref{eq:tr_estimate}) by solving Eqs (\ref{eq:kbvp_ode1})--(\ref{eq:kbvp_bc2}) with $k = 2$ and using the previously computed moment $u_{i,1}(r)$ \change{in the right-hand side of Eqs (\ref{eq:kbvp_ode1})--(\ref{eq:kbvp_ode3})}. The process repeats over $k$ until the value of $t_{r}^{\ast}$ converges sufficiently to an accurate estimate of $t_{r}$ as defined in Eq (\ref{eq:tr_definition}). \change{We remark that numerically solving the system (\ref{eq:kbvp_ode1})--(\ref{eq:kbvp_bc2}) is computationally inexpensive, since the coefficient matrix that arises from spatial discretisation of the boundary value problem remains unchanged throughout the iterations and only needs to be factorized once, with only the right-hand side of Eqs (\ref{eq:kbvp_ode1})--(\ref{eq:kbvp_ode3}) changing with $k$}.

\section{Results and discussion}
\label{sec:results}
In this section, we apply the characteristic timescale of Section \ref{sec:3} and the release time estimate of Section \ref{sec:4} to several test cases. For all our numerical experiments we consider the drug diffusion model (\ref{eq:model_pde1})--(\ref{eq:model_ic3}) with  $n = 1$ ({\em core-shell model}) and the following base values for the geometrical and 
physical parameters \cite{carr_2018mb,kaoui_2018, henning_2012}:
\begin{align}
\label{eq:par1}
&R_0=1.5 \cdot 10^{-3}\,\mathrm{m},\quad R_1=1.7 \cdot 10^{-3}\,\mathrm{m},\quad R_{\infty} = 30 \cdot 10^{-3}\,\mathrm{m},\quad \sigma_{0} = \sigma_{1} = 1,\\
&D_0=30 \cdot 10^{-11}\,\mathrm{m}^2\mathrm{s}^{-1},\quad D_1=5 \cdot 10^{-11}\,\mathrm{m}^2\mathrm{s}^{-1},\quad  D_e=30 \cdot 10^{-11}\,\mathrm{m}^2\,\mathrm{s}^{-1},\quad P \rightarrow \infty.
\label{eq:par2}
\end{align} 
Using this set of parameter values, we first investigate the convergence behaviour of the asymptotic estimate of the release time, $t_{r}^{\ast}$ (\ref{eq:tr_estimate}), for increasing values of $k$. In Table \ref{tab:convergence}, we report the value of $t_{r}^{\ast}$ \change{and $\widetilde{c}_{0}(t_{r}^{\ast})$} for $k = 1,\hdots,14$ and $\varepsilon = 10^{-4}, 10^{-5}, 10^{-6}$. These results show that $t_{r}^{\ast}$ is converging to a highly accurate estimate of the release time, with $\lim_{k\rightarrow\infty}\widetilde{c}_{0}(t_{r}^{\ast})$ close to the prescribed value of $\varepsilon$ in all three cases. \change{For the two smaller values of $\varepsilon$, $\lim_{k\rightarrow\infty}\widetilde{c}_{0}(t_{r}^{\ast})$ is closer to the specified value of $\varepsilon$ because the one-term approximation (\ref{eq:ci_asymptotic}) becomes more accurate for larger values of $t$}. For this parameter set, the release time estimate converges to the second by $k=12$ for all three values of $\varepsilon$.

\begin{table}[h]
\centering
\def\arraystretch{1.05}
\begin{tabular}{|c|rr|}
\hline
& \multicolumn{2}{c|}{$\varepsilon = 10^{-4}$}\\
\hline
$k$ & $t_{r}^{\ast}$ & $\widetilde{c}_{0}(t_{r}^{\ast})$\\
\hline
1 & 02:06:22:59 & $7.2152 \cdot 10^{-4}$\\
2 & 07:23:03:31 & $8.6792 \cdot 10^{-5}$\\
3 & 07:21:39:42 & $8.8261 \cdot 10^{-5}$\\
4 & 07:14:19:34 & $9.6411 \cdot 10^{-5}$\\
5 & 07:11:49:55 & $9.9360 \cdot 10^{-5}$\\
6 & 07:11:02:47 & $1.0031 \cdot 10^{-4}$\\
7 & 07:10:48:25 & $1.0060 \cdot 10^{-4}$\\
8 & 07:10:44:09 & $1.0069 \cdot 10^{-4}$\\
9 & 07:10:42:54 & $1.0071 \cdot 10^{-4}$\\
10 & 07:10:42:33 & $1.0072 \cdot 10^{-4}$\\
11 & 07:10:42:27 & $1.0072 \cdot 10^{-4}$\\
12 & 07:10:42:25 & $1.0072 \cdot 10^{-4}$\\
13 & 07:10:42:25 & $1.0072 \cdot 10^{-4}$\\
14 & 07:10:42:25 & $1.0072 \cdot 10^{-4}$\\
\hline
\end{tabular}%
\begin{tabular}{rr|}
\hline
\multicolumn{2}{c|}{$\varepsilon = 10^{-5}$}\\
\hline
$t_{r}^{\ast}$ & $\widetilde{c}_{0}(t_{r}^{\ast})$\\
\hline
02:22:53:05 & $4.8197 \cdot 10^{-4}$\\
13:02:30:33 & $2.0023 \cdot 10^{-5}$\\
15:11:32:11 & $1.0190 \cdot 10^{-5}$\\
15:14:06:10 & $9.8849 \cdot 10^{-6}$\\
15:13:37:58 & $9.9401 \cdot 10^{-6}$\\
15:13:18:53 & $9.9777 \cdot 10^{-6}$\\
15:13:11:17 & $9.9926 \cdot 10^{-6}$\\
15:13:08:41 & $9.9977 \cdot 10^{-6}$\\
15:13:07:52 & $9.9994 \cdot 10^{-6}$\\
15:13:07:37 & $9.9999 \cdot 10^{-6}$\\
15:13:07:32 & $1.0000 \cdot 10^{-5}$\\
15:13:07:31 & $1.0000 \cdot 10^{-5}$\\
15:13:07:31 & $1.0000 \cdot 10^{-5}$\\
15:13:07:31 & $1.0000 \cdot 10^{-5}$\\
\hline
\end{tabular}%
\begin{tabular}{rr|}
\hline
\multicolumn{2}{c|}{$\varepsilon = 10^{-6}$}\\
\hline
$t_{r}^{\ast}$ & $\widetilde{c}_{0}(t_{r}^{\ast})$\\
\hline
03:15:23:12 & $3.5016 \cdot 10^{-4}$\\
18:05:57:35 & $4.6400 \cdot 10^{-6}$\\
23:01:24:40 & $1.1822 \cdot 10^{-6}$\\
23:13:52:46 & $1.0199 \cdot 10^{-6}$\\
23:15:26:02 & $1.0013 \cdot 10^{-6}$\\
23:15:34:58 & $9.9953 \cdot 10^{-7}$\\
23:15:34:10 & $9.9969 \cdot 10^{-7}$\\
23:15:33:14 & $9.9988 \cdot 10^{-7}$\\
23:15:32:50 & $9.9996 \cdot 10^{-7}$\\
23:15:32:41 & $9.9999 \cdot 10^{-6}$\\
23:15:32:38 & $1.0000 \cdot 10^{-6}$\\
23:15:32:37 & $1.0000 \cdot 10^{-6}$\\
23:15:32:37 & $1.0000 \cdot 10^{-6}$\\
23:15:32:37 & $1.0000 \cdot 10^{-6}$\\
\hline
\end{tabular}
\caption{Convergence of the asymptotic estimate of the release time $t_{r}^{\ast}$ (\ref{eq:tr_estimate}) for increasing values of the index $k$ applied to the drug diffusion model (\ref{eq:model_pde1})--(\ref{eq:model_ic3}) with a single hydrogel layer ($n = 1$) and the physical parameters (\ref{eq:par1})--(\ref{eq:par2}). The value of $t_{r}^{\ast}$ is calculated using three different choices for $\varepsilon$ and is displayed in the format days:hours:minutes:seconds.}
\label{tab:convergence}
\end{table}

\begin{figure}[p]
\def\figureheight{0.25\textwidth}
\centering
\subfloat[(a)]{\includegraphics[height=\figureheight]{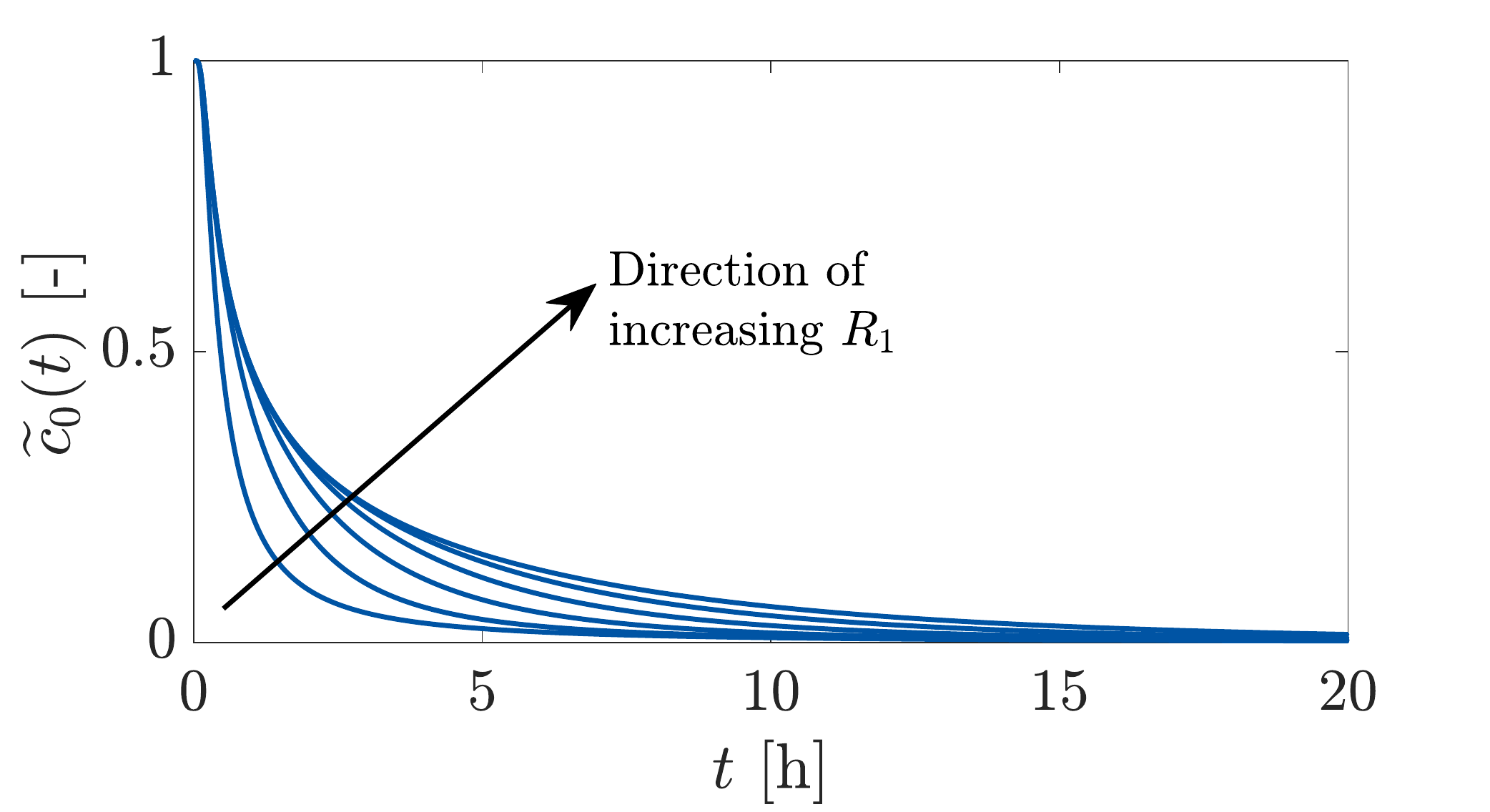}}\subfloat[(e)]{\includegraphics[height=\figureheight]{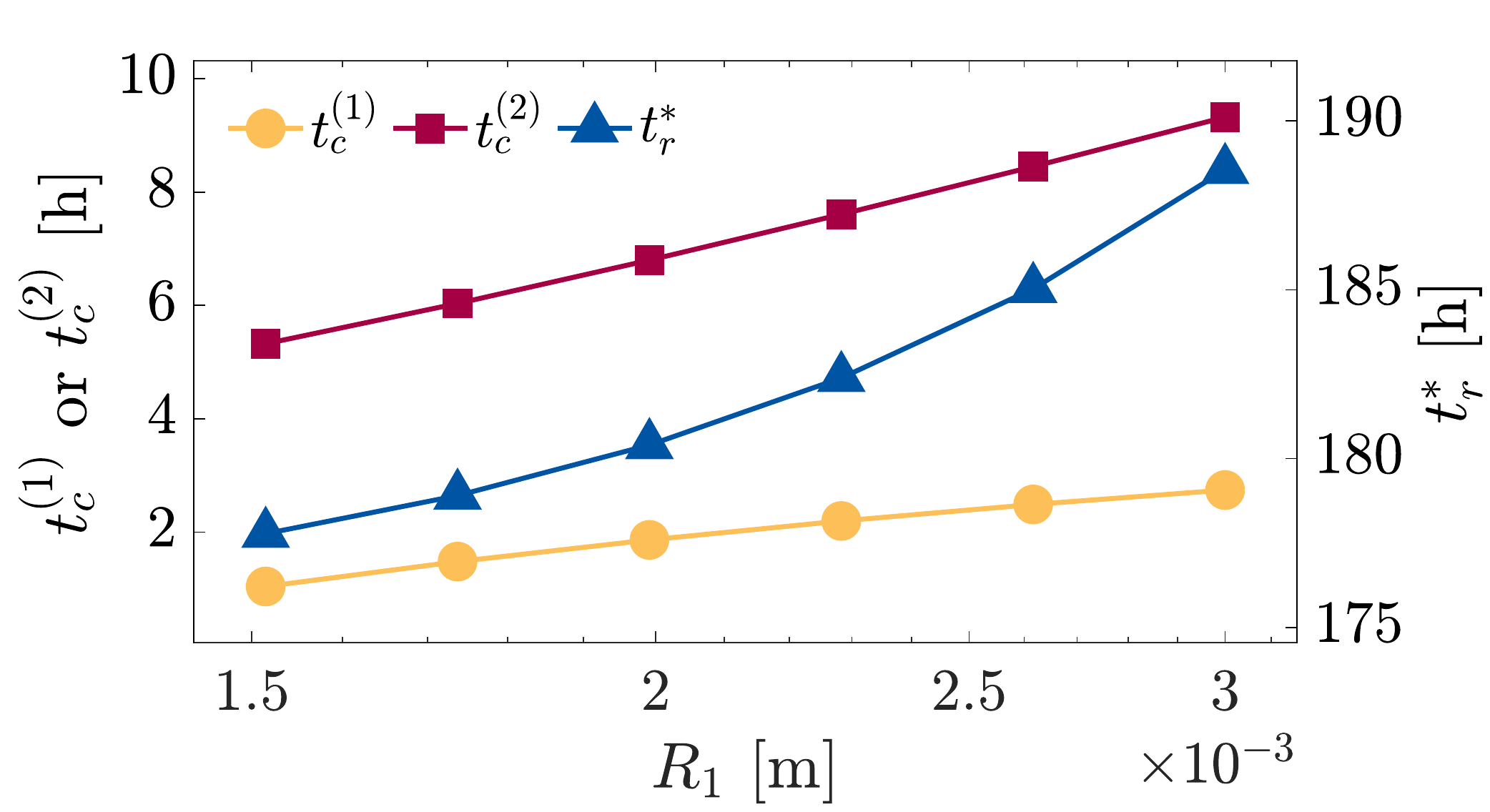}}\\
\subfloat[(b)]{\includegraphics[height=\figureheight]{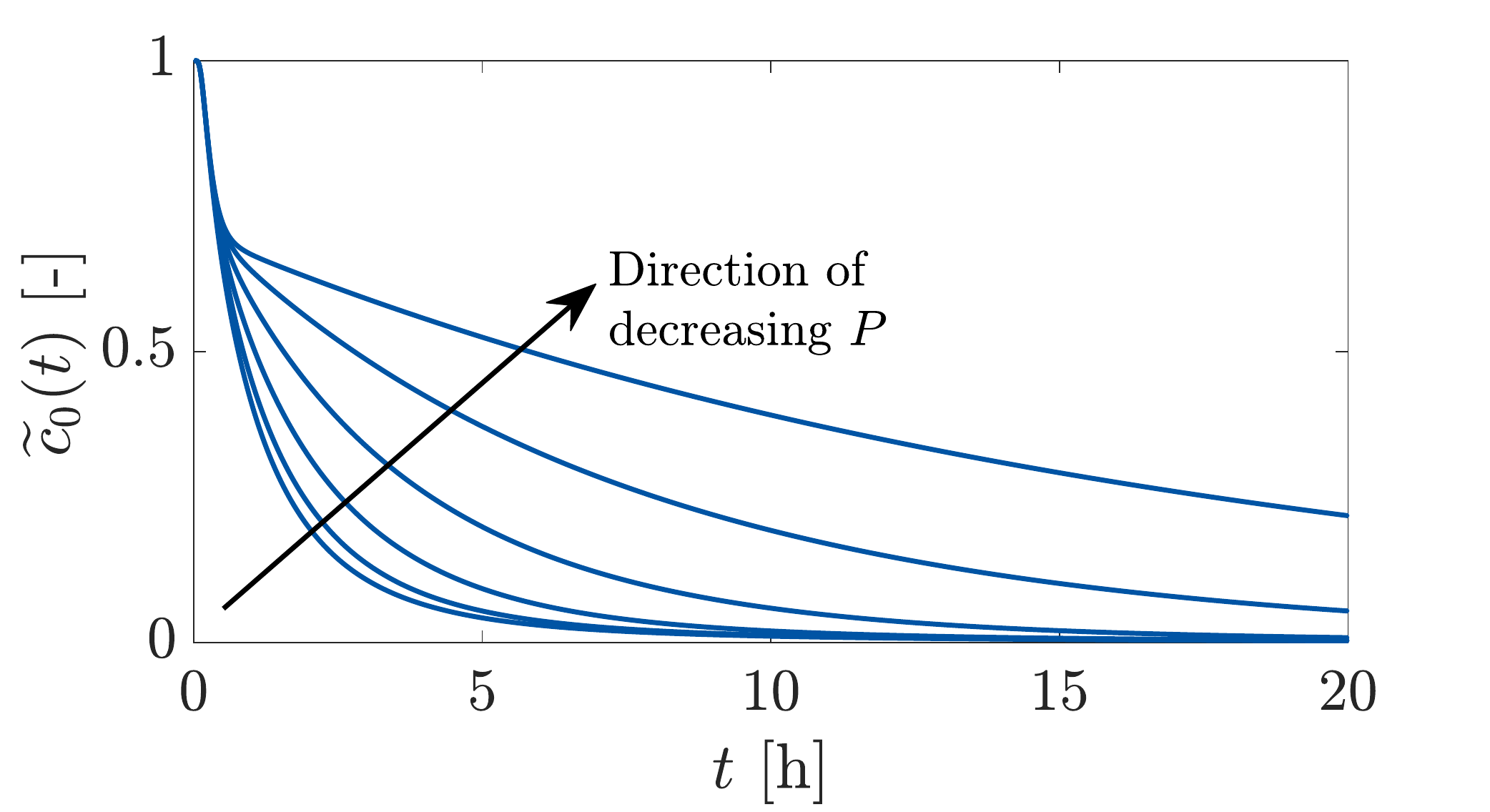}}\subfloat[(f)]{\includegraphics[height=\figureheight]{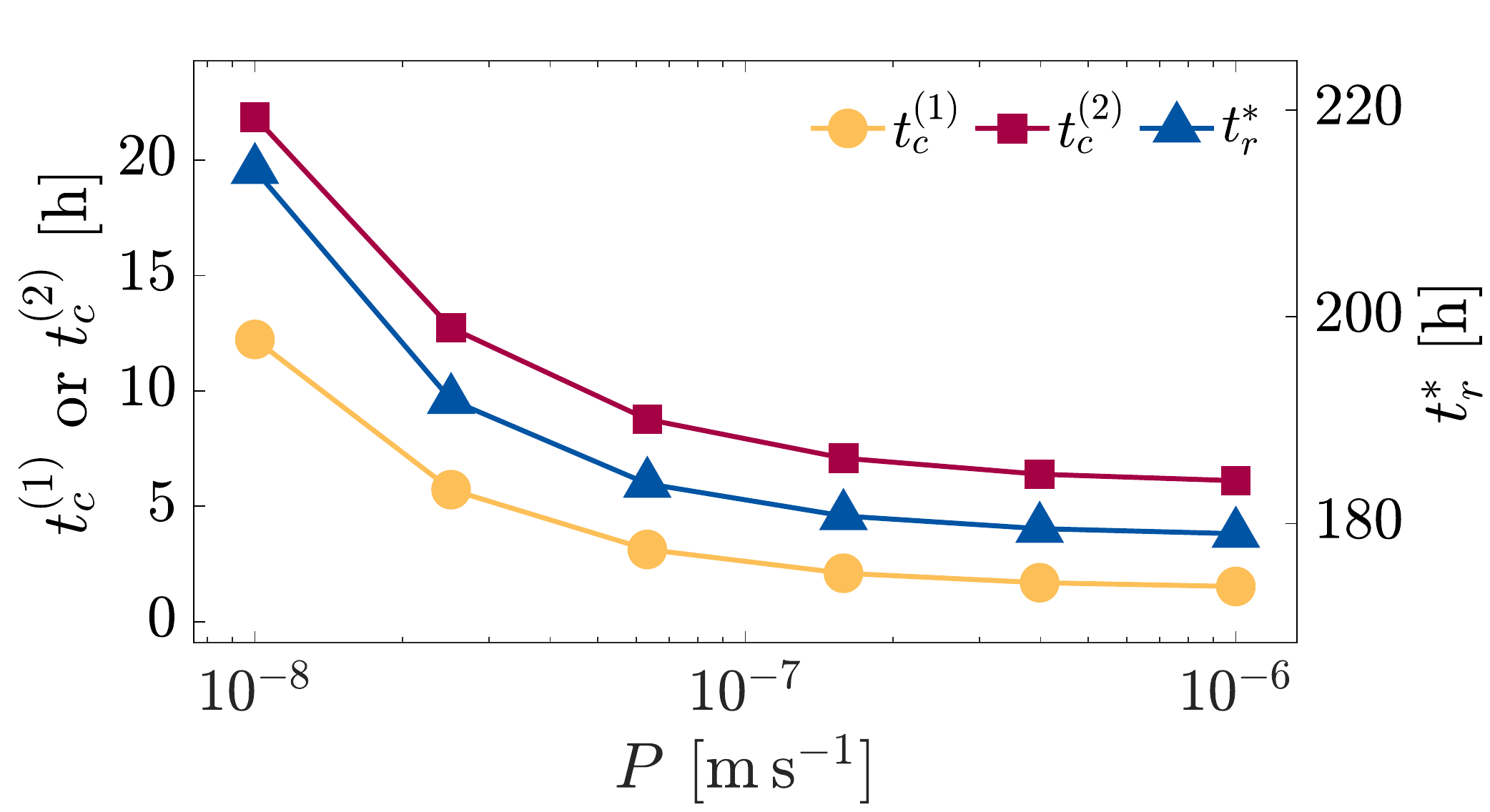}}\\
\subfloat[(c)]{\includegraphics[height=\figureheight]{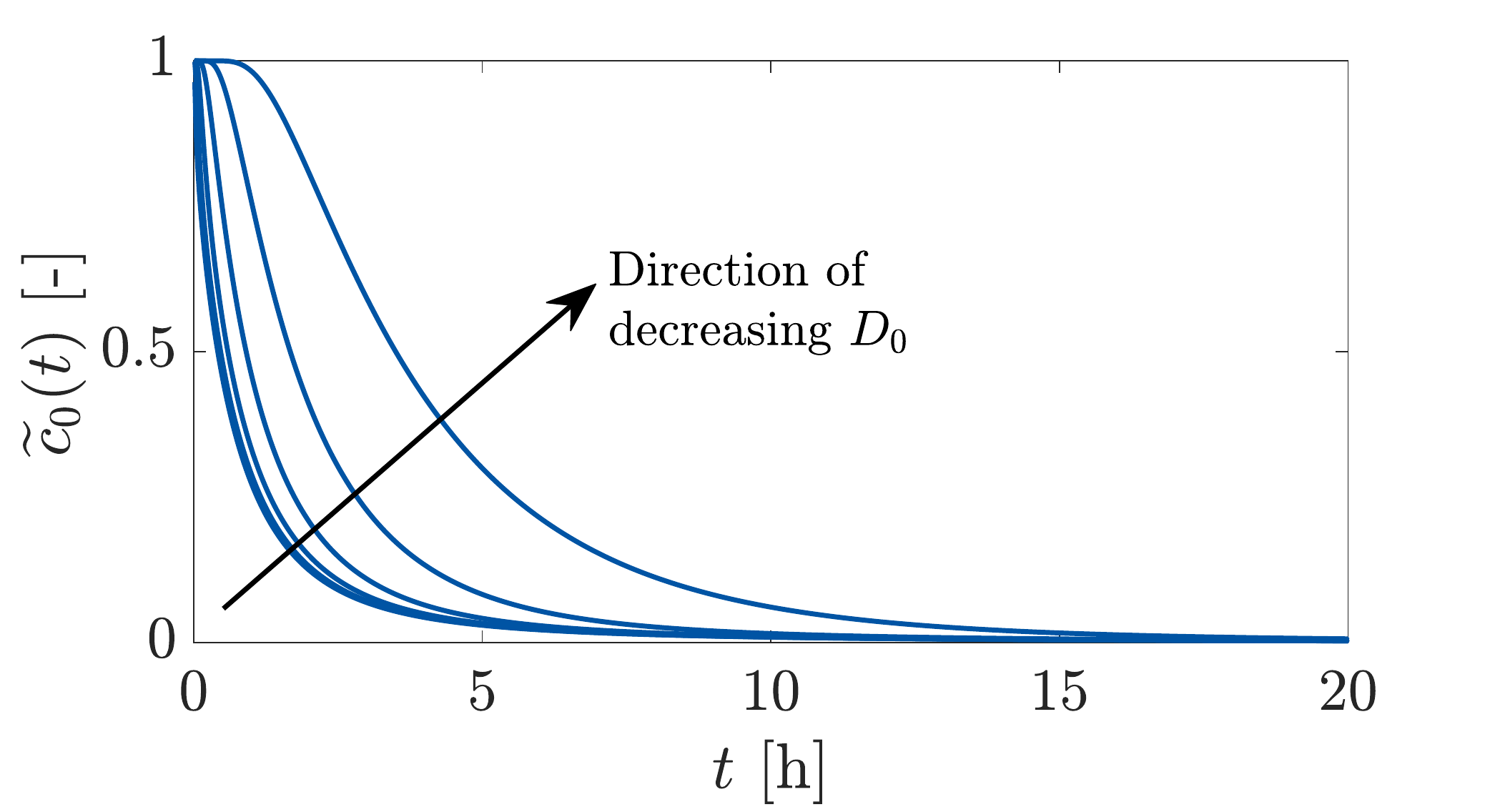}}\subfloat[(g)]{\includegraphics[height=\figureheight]{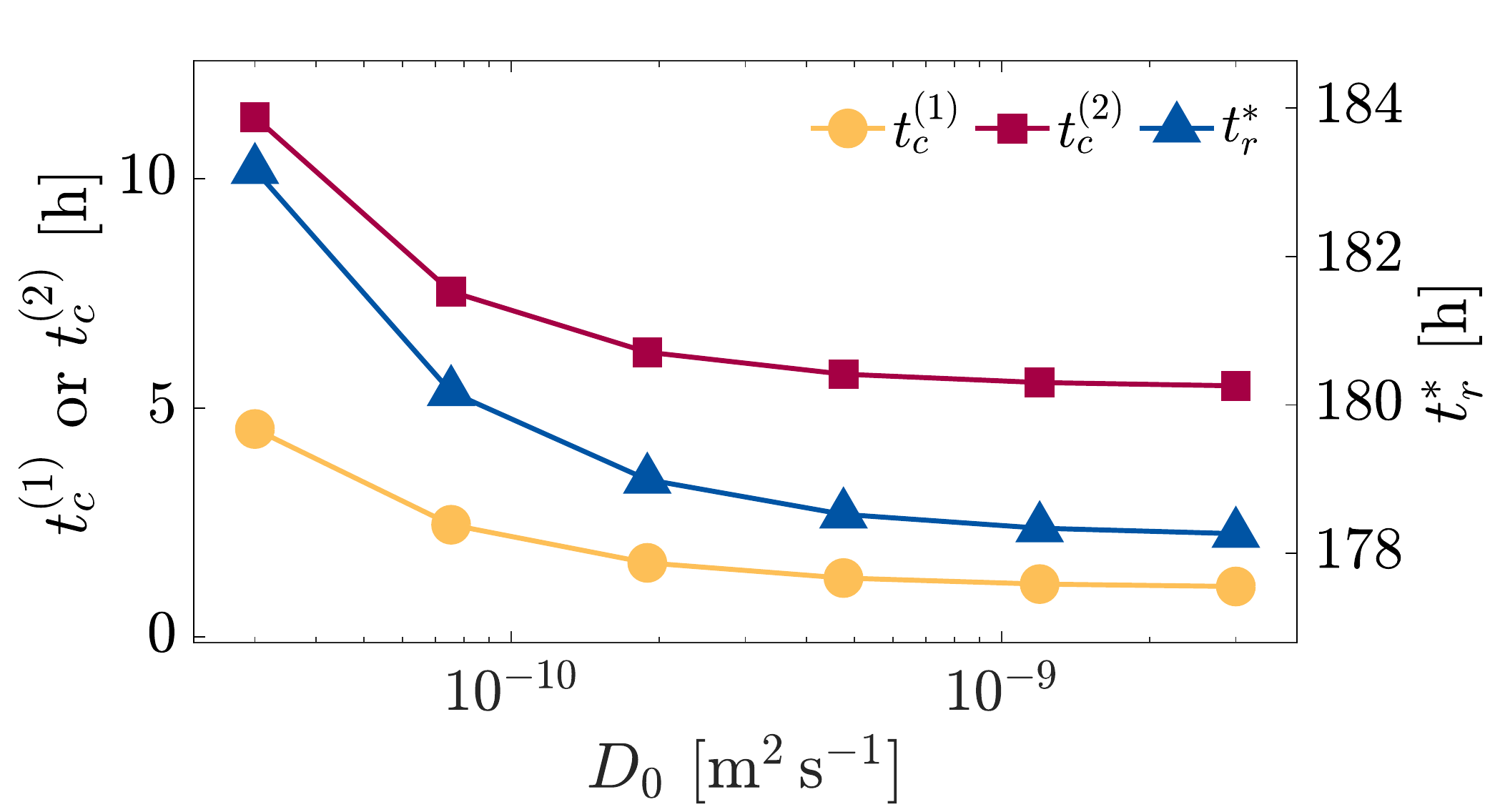}}\\
\subfloat[(d)]{\includegraphics[height=\figureheight]{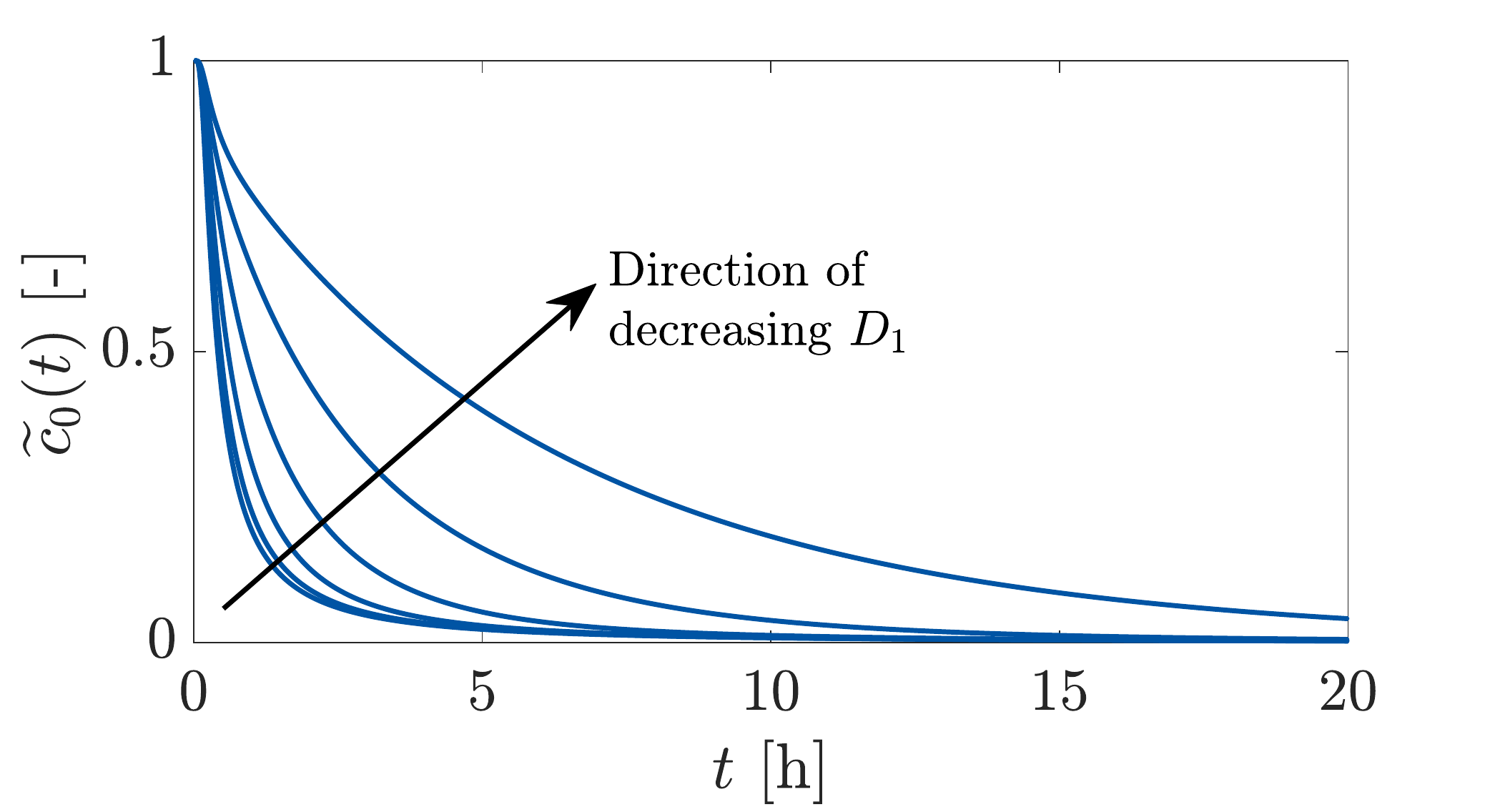}}\subfloat[(h)]{\includegraphics[height=\figureheight]{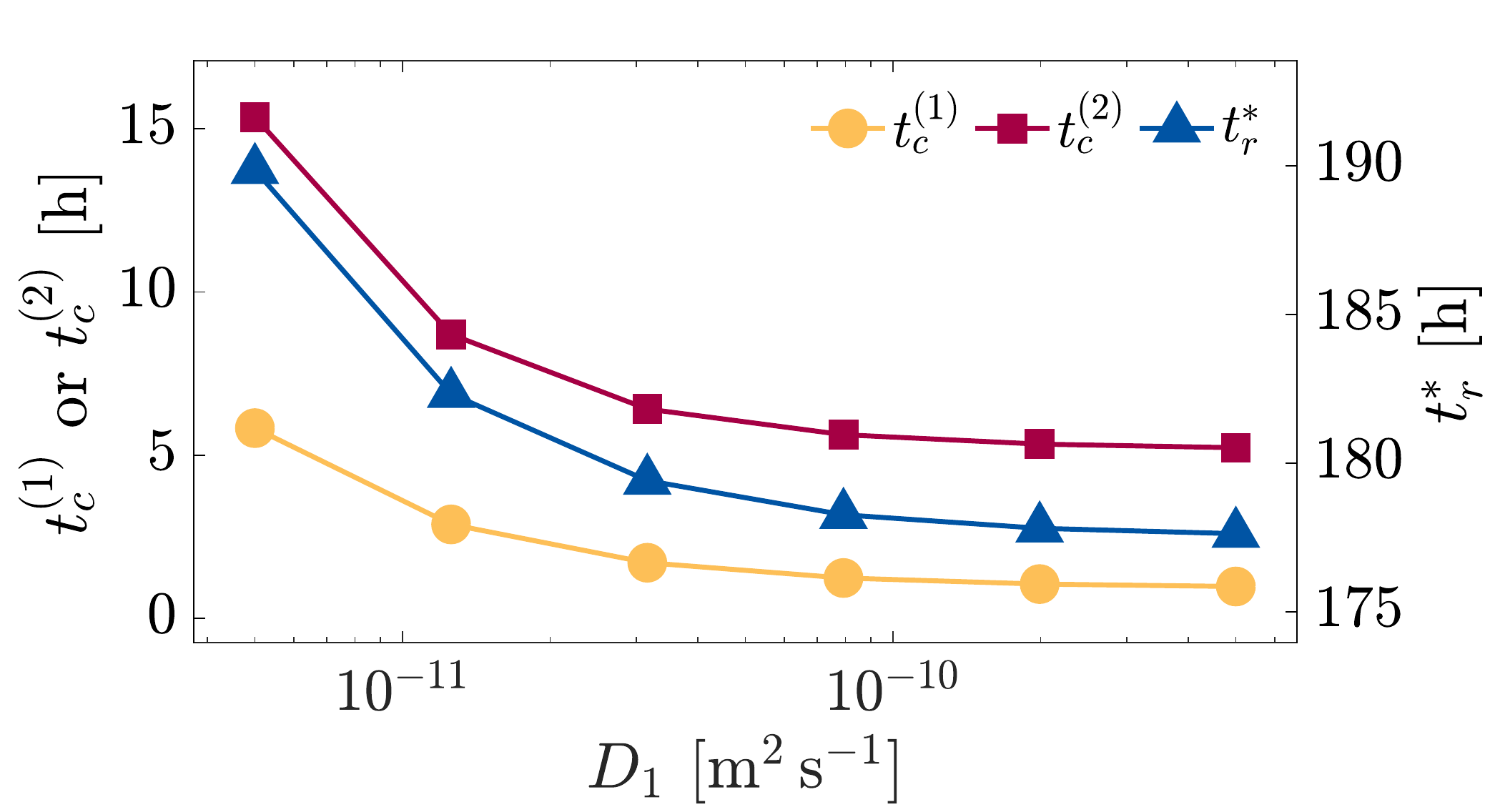}}\\
\caption{(a)--(d) Profiles of the dimensionless concentration at the centre of the capsule, $\widetilde{c}_{0}(t)$, over time showing the effect of varying (a) the outer radius of the capsule ($R_{1}$); (b) the mass transfer coefficient at the external coating ($P$); (c) the diffusivity in the core ($D_{0}$); and (d) the diffusivity in the hydrogel layer ($D_{1}$). (e)--(f) Sensitivity of the characteristic timescales, $t_{c}^{(1)}$ and $t_{c}^{(2)}$, and release time, $t_{r}^{\ast}$ \change{(calculated using $k = 15$ and $\varepsilon = 10^{-4}$)}, when varying (e) $R_{1}$; (f) $P$; (g) $D_{0}$; and (h) $D_{1}$. When varying a parameter all other variables are held fixed at the values given in Eqs (\ref{eq:par1})--(\ref{eq:par2}). All results are based on the drug diffusion model (\ref{eq:model_pde1})--(\ref{eq:model_ic3}) with a single hydrogel layer ($n = 1$) and the baseline parameter values in Eqs (\ref{eq:par1})--(\ref{eq:par2}) (cf.~Table \ref{tab:results}).}
\label{fig:results}
\end{figure} 

\begin{table}[t]
\centering
\def\arraystretch{1.05}
\begin{tabular}{|l|r|r|r|r|r|r|}
\hline
\multicolumn{7}{|l|}{\textit{Varying $R_{1}$}}\\
\hline
Case & 1 & 2 & 3 & 4 & 5 & 6\\
$R_{1}$ [m] & $1.52 \cdot 10^{-3}$ & $1.74 \cdot 10^{-3}$ & $1.99 \cdot 10^{-3}$ & $2.28 \cdot 10^{-3}$ & $2.62 \cdot 10^{-3}$ & $3.00 \cdot 10^{-3}$\\ 
$t_{c}^{(1)}$ [h] & $1.04$ & $1.48$ & $1.86$ & $2.20$ & $2.49$ & $2.74$\\ 
$t_{c}^{(2)}$ [h] & $5.32$ & $6.03$ & $6.80$ & $7.61$ & $8.45$ & $9.32$\\ 
$t_{r}^{\ast}$ [h] & $177.78$ & $178.91$ & $180.40$ & $182.39$ & $185.03$ & $188.55$\\ 
%$\widetilde{c}(t_{r}^{\ast})$ [-] & $1.01 \cdot 10^{-4}$ & $1.01 \cdot 10^{-4}$ & $1.01 \cdot 10^{-4}$ & $1.01 \cdot 10^{-4}$ & $1.01 \cdot 10^{-4}$ & $1.01 \cdot 10^{-4}$\\
\hline
\multicolumn{7}{|l|}{\textit{Varying $P$}}\\
\hline
Case & 7 & 8 & 9 & 10 & 11 & 12\\
$P$ [$\mathrm{m}\,\mathrm{s}^{-1}$] & $1.00 \cdot 10^{-8}$ & $2.51 \cdot 10^{-8}$ & $6.31 \cdot 10^{-8}$ & $1.58 \cdot 10^{-7}$ & $3.98 \cdot 10^{-7}$ & $1.00 \cdot 10^{-6}$\\ 
$t_{c}^{(1)}$ [h] & $12.23$ & $5.72$ & $3.13$ & $2.10$ & $1.69$ & $1.52$\\ 
$t_{c}^{(2)}$ [h] & $21.86$ & $12.73$ & $8.77$ & $7.08$ & $6.39$ & $6.11$\\ 
$t_{r}^{\ast}$ [h] & $214.19$ & $191.96$ & $183.86$ & $180.74$ & $179.51$ & $179.03$\\ 
%$\widetilde{c}(t_{r}^{\ast})$ [-] & $1.05 \cdot 10^{-4}$ & $1.01 \cdot 10^{-4}$ & $1.01 \cdot 10^{-4}$ & $1.01 \cdot 10^{-4}$ & $1.01 \cdot 10^{-4}$ & $1.01 \cdot 10^{-4}$\\
\hline
\multicolumn{7}{|l|}{\textit{Varying $D_{0}$}}\\
\hline
Case & 13 & 14 & 15 & 16 & 17 & 18\\
$D_{0}$ [$\mathrm{m}^{2}\,\mathrm{s}^{-1}$] & $3.00 \cdot 10^{-11}$ & $7.54 \cdot 10^{-11}$ & $1.89 \cdot 10^{-10}$ & $4.75 \cdot 10^{-10}$ & $1.19 \cdot 10^{-9}$ & $3.00 \cdot 10^{-9}$\\ 
$t_{c}^{(1)}$ [h] & $4.54$ & $2.45$ & $1.62$ & $1.29$ & $1.16$ & $1.10$\\ 
$t_{c}^{(2)}$ [h] & $11.35$ & $7.54$ & $6.22$ & $5.74$ & $5.55$ & $5.48$\\ 
$t_{r}^{\ast}$ [h] & $183.16$ & $180.17$ & $178.99$ & $178.53$ & $178.34$ & $178.27$\\ 
%$\widetilde{c}(t_{r}^{\ast})$ [-] & $1.01 \cdot 10^{-4}$ & $1.01 \cdot 10^{-4}$ & $1.01 \cdot 10^{-4}$ & $1.01 \cdot 10^{-4}$ & $1.01 \cdot 10^{-4}$ & $1.01 \cdot 10^{-4}$\\ 
\hline
\multicolumn{7}{|l|}{\textit{Varying $D_{1}$}}\\
\hline
Case & 19 & 20 & 21 & 22 & 23 & 24\\
$D_{1}$ [$\mathrm{m}^{2}\,\mathrm{s}^{-1}$] & $5.00 \cdot 10^{-12}$ & $1.26 \cdot 10^{-11}$ & $3.15 \cdot 10^{-11}$ & $7.92 \cdot 10^{-11}$ & $1.99 \cdot 10^{-10}$ & $5.00 \cdot 10^{-10}$\\ 
$t_{c}^{(1)}$ [h] & $5.83$ & $2.88$ & $1.70$ & $1.23$ & $1.05$ & $0.97$\\ 
$t_{c}^{(2)}$ [h] & $15.36$ & $8.70$ & $6.41$ & $5.63$ & $5.34$ & $5.23$\\ 
$t_{r}^{\ast}$ [h] & $189.86$ & $182.33$ & $179.41$ & $178.26$ & $177.81$ & $177.63$\\ 
%$\widetilde{c}(t_{r}^{\ast})$ [-] & $1.01 \cdot 10^{-4}$ & $1.01 \cdot 10^{-4}$ & $1.01 \cdot 10^{-4}$ & $1.01 \cdot 10^{-4}$ & $1.01 \cdot 10^{-4}$ & $1.01 \cdot 10^{-4}$\\ 
\hline\end{tabular}
\caption{Sensitivity of the characteristic timescales, $t_{c}^{(1)}$ and $t_{c}^{(2)}$, and release time, $t_{r}^{\ast}$ \change{with $\varepsilon = 10^{-4}$}, when varying the outer radius of the capsule ($R_{1}$); the mass transfer coefficient at the external coating ($P$); the diffusivity in the core ($D_{0}$); and the diffusivity in the hydrogel layer ($D_{1}$). When varying a parameter all other parameters are held fixed at the values given in Eqs (\ref{eq:par1})--(\ref{eq:par2}). All results are based on the drug diffusion model (\ref{eq:model_pde1})--(\ref{eq:model_ic3}) with a single hydrogel layer ($n = 1$). Values of $R_{1}$, $P$, $D_{0}$, $D_{1}$ are displayed to three significant figures while values of $t_{c}^{(1)}$, $t_{c}^{(2)}$ and $t_{r}^{\ast}$ are reported to two decimal places (cf.~Figure \ref{fig:results}).}
\label{tab:results}
\end{table}

We now investigate the sensitivity of $t_{r}^{\ast}$ when varying the outer radius of the capsule, $R_{1}$, the mass transfer coefficient at the external coating, $P$, the diffusivity in the core, $D_{0}$, and the diffusivity in the hydrogel layer, $D_{1}$, about the base values given in Eqs (\ref{eq:par1})--(\ref{eq:par2}). Each of the parameters, $R_{1}$, $P$, $D_{0}$ and $D_{1}$, are varied one at a time holding the other three parameters fixed at the values given in Eqs (\ref{eq:par1})--(\ref{eq:par2}). We consider six logarithmically-spaced values of $R_{1}$ between $1.515\cdot 10^{-3}$ and $3\cdot 10^{-3}$, $P$ between $10^{-8}$ and $10^{-6}$, $D_{0}$ between $3\cdot 10^{-11}$ and $3\cdot 10^{-9}$ and $D_{1}$ between $5\cdot 10^{-12}$ and $5\cdot 10^{-10}$. In total, there are 24 test cases as labelled in Table \ref{tab:results}. For all simulations, the release time estimate $t_{r}^{\ast}$ (\ref{eq:tr_estimate}) is calculated with $k = 15$ and $\varepsilon = 10^{-4}$. Results in Figures \ref{fig:results}(a)--(d) show the effect that varying each parameter has on the temporal profile of the dimensionless concentration at the centre of the capsule, $\widetilde{c}_{0}(t)$. These curves demonstrate that, over the specified ranges of each parameter, the time required for $\widetilde{c}_{0}(t)$ to approach zero increases for increasing values of $R_{1}$ and decreasing values of $P$, $D_{0}$ and $D_{1}$. This translates to larger values of $t_{r}^{\ast}$ for increasing values of $R_{1}$ and decreasing values of $P$, $D_{0}$ and $D_{1}$ \change{(holding all other parameters fixed)} as confirmed in Figures \ref{fig:results}(e)--(h) and Table \ref{tab:results}. 

An interesting exercise is to determine how well the characteristic timescales, $t_{c}^{(1)}$ and $t_{c}^{(2)}$, indicate the value of the release time. To explore this, we plot $t_{c}^{(1)}$ and $t_{c}^{(2)}$ versus $R_{1}$, $P$, $D_{0}$ and $D_{1}$ in Figure \ref{fig:results}(e)--(h) together with the corresponding values of the release time estimate $t_{r}^{\ast}$. The values of $t_{c}^{(1)}$ and $t_{c}^{(2)}$ are quite different from each other and differ considerably from $t_{r}^{\ast}$. However, this is not surprising because, as already mentioned, the goal of $t_{c}^{(1)}$ and $t_{c}^{(2)}$ is not to estimate the release time but rather provide a cheap way to rank different capsule configurations in terms of release time. Evident from these results is that both $t_{c}^{(1)}$ and $t_{c}^{(2)}$ follows the same general trend as $t_{r}^{\ast}$, that is, both increase with $R_{1}$ and decrease with $P$, $D_{0}$ and $D_{1}$. As the $\widetilde{c}_{0}(t)$ curves in Figures \ref{fig:results}(a)--(d) do not intersect, we are guaranteed that a smaller value of $t_{c}^{(1)}$ implies a smaller value of $t_{r}^{\ast}$, as confirmed in Figures \ref{fig:results}(e)--(h). An equivalent guarantee is true for $t_{c}^{(2)}$. Comparing across all 24 test cases listed in Table \ref{tab:results}, we see that Case 24 gives rise to both the smallest characteristic timescales and release time while Case 7 yields both the largest ones. 

\begin{figure}[h]
\centering
\includegraphics[height=0.27\textwidth]{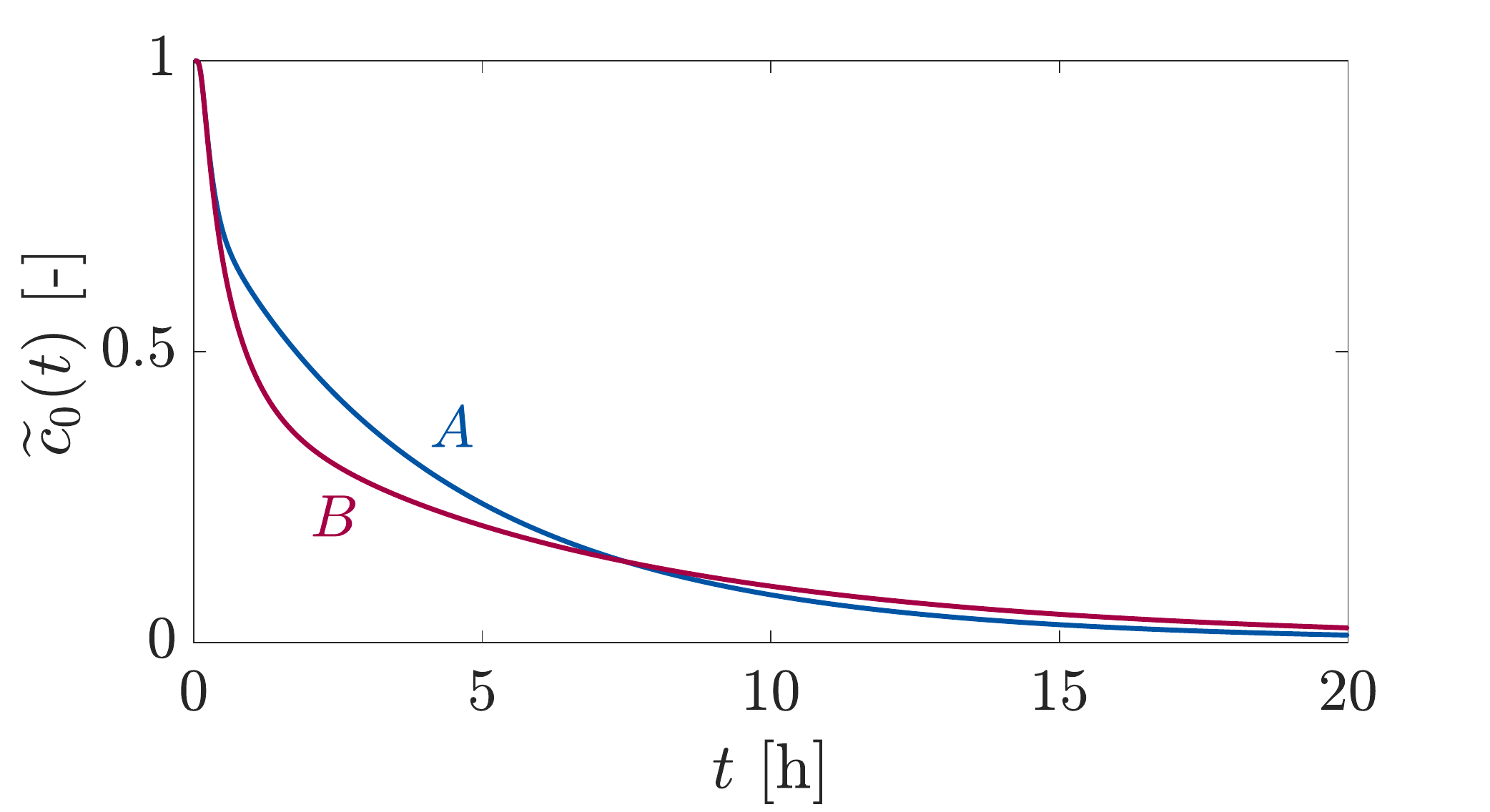}
\caption{Profile of the dimensionless concentration at the centre of the capsule, $\widetilde{c}_{0}(t)$, for two configurations of the drug diffusion model (\ref{eq:model_pde1})--(\ref{eq:model_ic3}), both with a single hydrogel layer ($n = 1$). Configuration $A$ has physical parameters (\ref{eq:par1})--(\ref{eq:par2}) except with $P = 5\cdot 10^{-8}\,\mathrm{m}\,\mathrm{s}^{-1}$ while configuration $B$ has physical parameters (\ref{eq:par1})--(\ref{eq:par2}) except with $P = 5\cdot 10^{-8}\,\mathrm{m}\,\mathrm{s}^{-1}$ and $R_{1} = 2.2\cdot 10^{-3}\,\mathrm{m}$. \change{For configuration $A$, $t_{c}^{(1)}=3.58\,\mathrm{h}$, $t_{c}^{(2)} = 9.48\,\mathrm{h}$ and $t_{r}^{\ast} = 185.23\,\mathrm{h}$ while for configuration $B$, $t_{c}^{(1)} = 3.40\,\mathrm{h}$, $t_{c}^{(2)} = 10.37\,\mathrm{h}$ and $t_{r}^{\ast} = 190.45\,\mathrm{h}$.}}
\label{fig:AB}
\end{figure}

The above results seem to indicate that the characteristic timescale always provides a useful indication of the release time. However, this is not necessarily true and caution is required \change{when using either $t_{c}^{(1)}$ (\ref{eq:tc1final}) or  $t_{c}^{(2)}$ (\ref{eq:tc2final})} as an indicator of the release time of different capsule configurations. For example, consider two configurations $A$ and $B$, where configuration $A$ has physical parameters (\ref{eq:par1})--(\ref{eq:par2}) except with $P = 5\cdot 10^{-8}\,\mathrm{m}\,\mathrm{s}^{-1}$ while configuration $B$ has physical parameters (\ref{eq:par1})--(\ref{eq:par2}) except with $P = 5\cdot 10^{-8}\,\mathrm{m}\,\mathrm{s}^{-1}$ and $R_{1} = 2.2\cdot 10^{-3}\,\mathrm{m}$. \change{In this case, the value of $t_{c}^{(1)}$ is larger for configuration $A$ despite the value of $t_{r}^{\ast}$ (calculated using $\varepsilon = 10^{-4}$ and $k = 15$) \change{being} smaller (see caption of Figure \ref{fig:AB} for the numerical values). In Figure \ref{fig:AB}, we also see that the dimensionless concentration at the centre of the capsule $\widetilde{c}_{0}(t)$ for configurations $A$ and $B$ intersect, meaning that a smaller value of $t_{c}^{(1)}$ is not guaranteed to produce a smaller release time (as pointed out in Section \ref{sec:3}). Conversely, the value of $t_{c}^{(2)}$ correctly correlates with $t_{r}^{\ast}$. Here, the second term in Eq (\ref{eq:tc2bis}) is larger for configuration $B$ due to its heavier tail, which causes the value of $t_{c}^{(2)}$ to be larger for configuration $B$ than for configuration $A$, consistent with $t_{r}^{\ast}$ (see caption of Figure \ref{fig:AB} for the numerical values)}.

\section{Conclusions} 
Polymer capsules, pellets, tablets, layer-by-layer vehicles and other micro-engineered drug releasing implants are attracting a great deal of attention for their potential use  for therapeutic applications. The design of these novel drug delivery systems \change{poses} major challenges, such as the unknown significance of process parameters. Specifically, modelling and computational tools are provided to assist in the development of drug-release devices that can control the time to establish a steady-state flux or to deliver at a desired rate. The performance of a composite microcapsule can be sensibly enhanced if the release mechanism is understood and an appropriate mathematical model is used to characterize the releasing ability of the system.  In this study, \change{novel approaches to characterize and estimate} the release time of diffusion problems from spherical multi-layer capsules are proposed under a limited number of physical assumptions. The method is based on the linearity of a pure diffusive system, that holds or dominates in most circumstances. No explicit solution of the diffusion problem is required\change{; instead temporal moments of the drug concentration versus time curve at the centre of the capsule are used} to derive analytical expressions that \change{provide \textit{a priori}} quantitative indication of the drug release time. The proposed methodology, \change{which combines} in just one quantity the relevant geometrical and physical parameters, provides a simple tool to measure microcapsule dynamic performance.

\section*{Acknowledgments}
G.P. acknowledges funding from
the European Research Council under the European Unions Horizon 2020 Framework Programme
(No.~FP/2014-2020)/ERC Grant Agreement No.~739964 (COPMAT).

\bibliographystyle{plainnat}

\end{document}